\documentclass[11pt, draftcls, onecolumn]{IEEEtran}

\usepackage{amsmath,amssymb,mathrsfs,graphicx}
\usepackage{psfrag}
\usepackage{nccmath}
\usepackage[ulem=normalem]{changes}

\let\oldbrace\{
\def\{{\oldbrace\kern0.5pt}

\def\tr{\mathop{\rm tr}\nolimits}%
\newcommand{\Ac}{\mathcal{A}}

\newcommand{\Cc}{\mathcal{C}}

\newcommand{\Dc}{\mathcal{D}}
\newcommand{\Ec}{\mathcal{E}}

\newcommand{\Gc}{\mathcal{G}}

\newcommand{\Kc}{\mathcal{K}}

\newcommand{\Nc}{\mathcal{N}}

\newcommand{\Sc}{\mathcal{S}}
\newcommand{\Tc}{\mathcal{T}}

\newcommand{\Xc}{\mathcal{X}}
\newcommand{\Yc}{\mathcal{Y}}

\newcommand{\Lv}{\mathbf{L}}
\newcommand{\Xv}{\mathbf{X}}
\newcommand{\Yv}{\mathbf{Y}}

\newcommand{\Uv}{\mathbf{U}}

\newcommand{\lv}{\mathbf{l}}
\newcommand{\mv}{\mathbf{m}}
\newcommand{\xv}{\mathbf{x}}
\newcommand{\yv}{\mathbf{y}}

\newcommand{\uv}{\mathbf{u}}

\newcommand{\pen}{{P_e^{(n)}}}

\newcommand{\aep}{{\mathcal{T}_{\epsilon}^{(n)}}}
\newcommand{\aepvar}{{\mathcal{T}_{\epsilon'}^{(n)}}}

\newcommand{\Mh}{{\hat{M}}}
\newcommand{\Rh}{{\hat{R}}}

\newcommand{\Yh}{{\hat{Y}}}
\newcommand{\Zh}{{\hat{Z}}}

\newcommand{\lh}{{\hat{l}}}
\newcommand{\mh}{{\hat{m}}}

\newcommand{\yh}{{\hat{y}}}

\newcommand{\Yvh}{\hat{\mathbf{Y}}}

\newcommand{\yvh}{\hat{\mathbf{y}}}

\newcommand{\Yt}{{\tilde{Y}}}

\newcommand{\yt}{{\tilde{y}}}

\def\e{\epsilon}

\DeclareMathOperator\E{\sf E}
\let\P\relax
\DeclareMathOperator\P{\sf P}

\DeclareMathOperator\C{C}

\newcommand{\N}{\mathrm{N}}

\newcommand{\Real}{\mathbb{R}}

\newtheorem{theorem}{Theorem}

\newtheorem{corollary}{Corollary}

\begin{document}
\title{Noisy Network Coding}

\author{Sung Hoon Lim, Young-Han Kim, Abbas El Gamal, and Sae-Young Chung%
\thanks{S.~H.~Lim and S.-Y.~Chung are with the Department of Electrical Engineering, KAIST,
  Daejeon 305-701, Korea (e-mail: sunghlim@kaist.ac.kr and
  sychung@ee.kaist.ac.kr).}%

\thanks{Y.-H.~Kim is with the Department of Electrical and Computer
  Engineering, University of California, San Diego, La Jolla, CA 92093,
  USA (e-mail: yhk@ucsd.edu).}
 \thanks{A.~El~Gamal is with the Department of Electrical Engineering,
Stanford University, Stanford, CA 94305, USA (e-mail: abbas@ee.stanford.edu).}}
\maketitle

\IEEEpeerreviewmaketitle

\begin{abstract}
A noisy network coding scheme for sending multiple sources over a
general noisy network is presented. For multi-source multicast
networks, the scheme naturally extends both network coding over
noiseless networks by Ahlswede, Cai, Li, and Yeung, and
compress--forward coding for the relay channel by Cover and El Gamal
to general discrete memoryless and Gaussian networks. The scheme also
recovers as special cases the results on coding for wireless relay
networks and deterministic networks by Avestimehr, Diggavi, and Tse,
and coding for wireless erasure networks by Dana, Gowaikar, Palanki,
Hassibi, and Effros. The scheme involves message repetition coding,
relay signal compression, and simultaneous decoding. Unlike previous
compress--forward schemes, where independent messages are sent over
multiple blocks, the same message is sent multiple times using
independent codebooks as in the network coding scheme for cyclic
networks. Furthermore, the relays do not use Wyner--Ziv binning as in
previous compress--forward schemes, and each decoder performs
simultaneous joint typicality decoding on the received signals from
all the blocks without explicitly decoding the compression indices. A
consequence of this new scheme is that achievability is proved simply
and more generally without resorting to time expansion to extend
results for acyclic networks to networks with cycles.  The noisy
network coding scheme is then extended to general multi-source
networks by combining it with decoding techniques for interference
channels. For the Gaussian multicast network, noisy network coding
improves the previously established gap to the cutset bound.  We also
demonstrate through two popular AWGN network examples that noisy
network coding can outperform conventional compress--forward,
amplify--forward, and hash--forward coding schemes.
\end{abstract}

\clearpage
\section{Introduction} \label{sec:introduction}

\begin{figure}[h]
\begin{center}
\small
\psfrag{p}[c]{$p(y_1,\ldots,y_N|x_1,x_2,\ldots,x_N)$}
\psfrag{x1}[r]{$M_1 \to (X_1,Y_1)$\hspace{-6pt}}
\psfrag{w1}[r]{$M$}
\psfrag{x2}[cl]{\hspace{3pt}$(X_N,Y_N) \leftarrow M_N$}
\psfrag{w2}[b]{$\Mh_N$}
\psfrag{xn}[tr]{$M_2 \to (X_2,Y_2)$\hspace{-20pt}}
\psfrag{wn}[r]{$M_{2k_2}$}
\psfrag{xk}[bc]{$M_k \to (X_k,Y_k)$\qquad}
\psfrag{net}[c]{}
\psfrag{wk}[b]{$\Mh_j$}\includegraphics[width=0.4\textwidth]{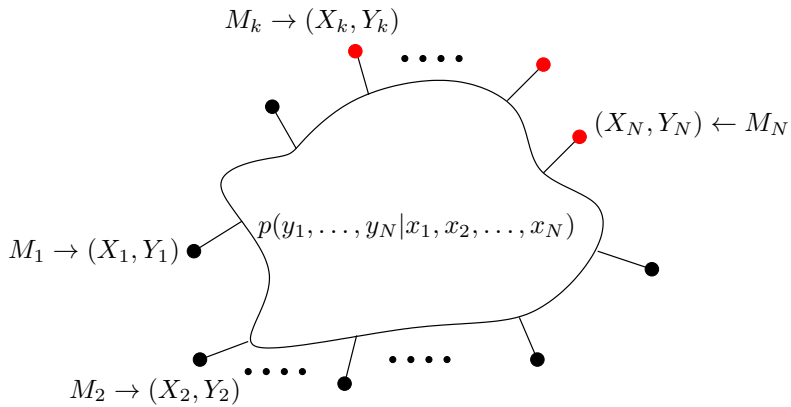}
\end{center}
\caption{An $N$-node discrete memoryless network.}
\label{fig:dmn}
\end{figure}

Consider the $N$-node discrete memoryless network depicted in
Figure~\ref{fig:dmn}. Each node wishes to send a message to a set of
destination nodes while acting as a relay for messages from other
nodes. What is the capacity region of this network, that is, the set
of rates at which the nodes can reliably communicate their messages?
What is the coding scheme that achieves the capacity region?  These
questions are at the heart of network information theory, yet complete
answers remain elusive.

Some progress has been made toward answering these questions in the
past forty years.  In~\cite{El-Gamal1981b, Cover--Thomas2006}, a
general cutset outer bound on the capacity region of this network was
established. This bound generalizes the max-flow min-cut theorem for
noiseless single-source unicast networks~\cite{Ford--Fulkerson1956,
  Elias--Feinstein--Shannon1956}, and has been shown to be tight for
several other classes of networks.

In their seminal paper on network
coding~\cite{Ahlswede--Cai--Li--Yeung2000}, Ahlswede, Cai, Li, and
Yeung showed that the capacity of noiseless single-source multicast
networks coincides with the cutset bound, thus generalizing the
max-flow min-cut theorem to multiple destinations. Each relay in the
network coding scheme sends a function of its incoming signals over
each outgoing link instead of simply forwarding incoming
signals. Their proof of the {\em network coding theorem} is done in
two steps. For acyclic networks, relay mappings are randomly generated
and they show that the message is correctly decoded with high
probability provided the rate is below the cutset bound. This proof is
then extended to cyclic networks by constructing an acyclic {\em
  time-expanded} network and relating achievable rates and codes for
the time-expanded network to those for the original cyclic network.

The network coding theorem has been extended in several
directions. Dana, Gowaikar, Palanki, Hassibi, and
Effros~\cite{Dana--Gowaikar--Palanki--Hassibi--Effros2006} studied the
multiple-source multicast erasure network as a simple model for a
wireless data network with packet loss. They showed that the capacity
region coincides with the cutset bound and is achieved via network
coding.  Ratnakar and Kramer~\cite{Ratnakar--Kramer2006} extended
network coding to characterize the multicast capacity for
single-source deterministic networks with broadcast but no
interference at the receivers.  Avestimehr, Diggavi, and Tse
\cite{Avestimehr--Diggavi--Tse2009} further extended this result to
deterministic networks with broadcast \emph{and} interference to
obtain a lower bound on capacity that coincides with the cutset bound
when the channel output is a linear function of input signals over a
finite field.  Their proof is again done in two steps. As in the
original proof of the network coding theorem, random coding is used to
establish the lower bound for {\em layered} deterministic networks. A
time-expansion technique is then used to extend the capacity lower
bound to arbitrary nonlayered deterministic networks.

In an earlier and seemingly unrelated line of investigation, van der
Meulen~\cite{van-der-Meulen1971a} introduced the relay channel with a
single source $X_1$, single destination $Y_3$, and single relay with
transmitter--receiver pair $(X_2,Y_2)$. Although the capacity for this
channel is still not known in general, several nontrivial upper and
lower bounds have been developed. In~\cite{Cover--El-Gamal1979}, Cover
and El Gamal proposed the compress--forward coding scheme in which the
relay compresses its noisy observation of the source signal and
forwards the compressed description to the destination.  Despite its
simplicity, compress--forward was shown to be optimal for classes of
deterministic~\cite{Kim2008a} and
modulo-sum~\cite{Aleksic--Razaghi--Yu2009} relay channels.  The
Cover--El Gamal compress--forward lower bound on capacity has the form
\begin{equation}
\label{eq:ce-cf}
C \ge \max_{p(x_1)p(x_2)p(\yh_2|y_2,x_2)} I(X_1;\Yh_2,Y_3|X_2),
\end{equation}
where the maximum is over all pmfs $p(x_1)p(x_2)p(\yh_2|y_2,x_2)$ such
that $I(X_2;Y_3) \ge I(Y_2;\Yh_2|X_2,Y_3)$. This lower bound was
established using a block Markov coding scheme---in each block the
sender transmits a new message, and the relay compresses its received
signal and sends the bin index of the compression index to the
receiver using Wyner--Ziv coding~\cite{Wyner--Ziv1976}. Decoding is
performed sequentially. At the end of each block, the receiver first
decodes the compression index and then uses it to decode the message
sent in the previous block. Kramer, Gastpar, and Gupta
\cite{Kramer--Gastpar--Gupta2005} used an extension of this scheme to
establish a compress--forward lower bound on the capacity of general
relay networks. Around the same time, El~Gamal, Mohseni, and
Zahedi~\cite{El-Gamal--Mohseni--Zahedi2006} put forth the equivalent
characterization of the compress--forward lower bound
\begin{align}
C &\ge \max_{p(x_1)p(x_2)p(\hat{y}_2|y_2,x_2)}
\min \{I(X_1;\Yh_2,Y_3|X_2),\, I(X_1,X_2;Y_3)-I(Y_2;\hat{Y}_2|X_1,X_2,Y_3)\}.\label{eq:emz-cf}
\end{align}
As we will see, this characterization motivates a more general way to
extend compress--forward to networks.

In this paper, we describe a noisy network coding scheme that extends
and unifies the above results. On the one hand, the scheme naturally
extends compress--forward coding to noisy networks. The resulting
inner bound on the capacity region extends the equivalent
characterization in~(\ref{eq:emz-cf}), rather than the original
characterization in~(\ref{eq:ce-cf}).  On the other hand, our scheme
includes network coding and its variants as special cases.  Hence,
while the coding schemes for deterministic networks and erasure
networks can be viewed as bottom-up generalizations of network coding
to more complicated networks, our coding scheme represents a top-down
approach for general noisy networks.

The noisy coding scheme employs block Markov message repetition
coding, relay signal compression, and simultaneous decoding.  Instead
of sending different messages over multiple blocks and decoding one
message at a time as in previous compress--forward coding
schemes~\cite{Cover--El-Gamal1979, Kramer--Gastpar--Gupta2005}, the
source transmits the same message over multiple blocks using
independently generated codebooks.  Although a similar message
repetition scheme is implicitly used in the time expansion technique
for cyclic noiseless networks~\cite{Ahlswede--Cai--Li--Yeung2000} and
nonlayered deterministic networks~\cite{Avestimehr--Diggavi--Tse2009},
our achievability proof does not require a two-step approach that
depends on the network topology.  The relay operation is also simpler
than previous compress--forward schemes---the compression index of the
received signal in each block is sent without Wyner--Ziv binning.
After receiving the signals from all the blocks, each destination node
performs simultaneous joint typicality decoding of the messages
without explicitly decoding the compression indices. As we will
demonstrate, this results in better performance than previous schemes
in~\cite{Kramer--Gastpar--Gupta2005, Rankov--Wittenben2006,
  Katti--Maric--Goldsmith--Katabi--Medard2007,
  Djeumou--Belmaga--Lasaulce2009, Razaghi--Yu2010} for networks with
more than one relay node or multiple messages.

The simplicity of our scheme makes it straightforward to combine with
decoding techniques for interference channels.  Indeed, the noisy
network coding scheme can be viewed as transforming a multi-hop relay
network into a single-hop interference network where the channel
outputs are compressed versions of the received signals.  We develop
two coding schemes for general multiple source networks based on this
observation. At one extreme, noisy network coding is combined with
decoding all messages, while at the other, interference is treated as
noise.

We apply these noisy network coding schemes to Gaussian networks. For
the multiple-source multicast case, we establish an inner bound that
improves upon previous capacity approximation results by Avestimehr,
Diggavi, and Tse~\cite{Avestimehr--Diggavi--Tse2009} and
Perron~\cite{Perron2009} with a tighter gap to the cutset bound.  We
then show that noisy network coding can outperform other specialized
schemes for two-way relay channels~\cite{Rankov--Wittenben2006,
  Katti--Maric--Goldsmith--Katabi--Medard2007} and interference relay
channels~\cite{Djeumou--Belmaga--Lasaulce2009, Razaghi--Yu2010}.

The rest of the paper is organized as follows. In the next section, we
formally define the problem of communicating multiple sources over a
general network and discuss the main results. We also show that previous
results on network coding are special cases of our main theorems and
compare noisy network coding to other schemes.  In
Section~\ref{sec:mmn}, we present the noisy network coding scheme for
multiple-source multicast networks. In Section~\ref{sec:general}, the
scheme is extended to general multiple-source networks.  Results on
Gaussian networks are discussed in Section~\ref{sec:gaussian}.

Throughout the paper, we follow the notation in
\cite{El-Gamal--Kim2010}.  In particular, a sequence of random
variables with node index $k$ and time index $i \in [1:n]$ is denoted
as $X_k^n=(X_{k1},\ldots, X_{kn})$. A set of random variables is
denoted as $X(\Ac)=\{X_k: k\in \Ac\}$.


\section{Problem Setup and Main Results} \label{sec:prob-statement}

The $N$-node discrete memoryless network (DMN) $(\prod_{k=1}^N \Xc_k,
p(y^N|x^N), \prod_{k=1}^N \Yc_k)$ depicted in Figure~\ref{fig:dmn}
consists of $N$ sender--receiver alphabet pairs $(\Xc_k, \Yc_k)$, $k
\in [1:N] := \{1, \ldots, N\}$, and a collection of conditional pmfs
$p(y_1,\ldots,y_N| x_1,\ldots,x_N)$. Each node $k \in [1:N]$ wishes to
send a message $M_k$ to a set of destination nodes, $\Dc_k \subseteq
[1:N]$.  Formally, a $(2^{nR_1}, \ldots, 2^{nR_N}, n)$ code for a DMN
consists of $N$ message sets $[1:2^{nR_1}], \ldots, [1:2^{nR_N}]$, a
set of encoders with encoder $k\in [1:N]$ that assigns an input symbol
$x_{ki}$ to each pair $(m_k, y^{i-1}_k)$ for $i\in[1:n]$, and a set of
decoders with decoder $d\in \cup_{k=1}^N \Dc_k$ that assigns message
estimates $(\mh_{kd}: k\in \Sc_d)$ to each $(y_d^n, m_d)$, where
$\Sc_d:=\{k: d\in \Dc_k\}$ is the set of nodes that send messages to
destination $d$. For simplicity we assume $d\in \Sc_d$ for all
destination nodes.

We assume that the messages $M_k$, $k\in [1:N]$, are independent of
each other and each message is uniformly distributed over its message
set. The average probability of error is defined as
\[
\pen= \P\{ \Mh_{kd}\neq M_{k} \text{ for some } d \in \Dc_k, k\in[1:N]\}.
\]
A rate tuple $(R_1, \ldots, R_N)$ is said to be achievable if there
exists a sequence of $(2^{nR_1}, \ldots, 2^{nR_N}, n)$ codes with
$\pen \to 0$ as $n\to\infty$.  The capacity region of the DMN is the
closure of the set of achievable rate tuples. 

We are ready to state our main results.
\medskip

\noindent{\bf Multiple-source multicast networks:} In
Section~\ref{sec:mmn}, we establish the following noisy network coding
theorem for multicasting multiple sources over a DMN. The coding
scheme and techniques used to prove this theorem, which we highlighted
earlier, constitute the key contributions of our paper.

\begin{theorem} \label{thm:mmn}
Let $\Dc = \Dc_1 = \cdots = \Dc_N$.  A rate tuple $(R_1, \ldots, R_N)$
is achievable for the DMN $p(y^N|x^N)$ if there exists some joint pmf
$p(q)\prod_{k=1}^{N}p(x_k|q) p(\yh_k|y_k,x_k,q)$ such that
\begin{equation}
\label{eq:mmn}
R(\Sc) < \min_{d\in \Sc^c \cap \Dc} I(X(\Sc); \Yh (\Sc^c),
Y_d|X(\Sc^c), Q) - I(Y(\Sc);\Yh(\Sc)|X^{N}, \Yh(\Sc^c), Y_d, Q)
\end{equation}
for all {\em cutsets} $\Sc\subseteq [1:N]$ with $\Sc^c \cap \Dc \neq
\emptyset$, where $R(\Sc)=\sum_{k\in \Sc} R_k$.
\end{theorem}

This inner bound has a similar structure to the cutset outer bound given by
\begin{equation} \label{eq:cutset-mmn}
R(\Sc) \leq  I(X(\Sc); Y(\Sc^c)| X(\Sc^c))
\end{equation}
for all $\Sc\subseteq [1:N]$ with $\Sc^c\cap \Dc\neq \emptyset$.  The
first term of \eqref{eq:cutset-mmn}, however, has $Y$ replaced by the
``compressed'' version $\Yh$. Another difference between the bounds is
the negative term appearing in \eqref{eq:mmn}, which quantifies
the rate requirement to convey the compressed version. In addition,
the maximum in \eqref{eq:mmn} is only over independent $X^N$.

Theorem~\ref{thm:mmn} can be specialized to several important
network models as follows:

\noindent {\em Noiseless networks:} Consider a noiseless network
modeled by a weighted directed graph $\Gc=(\Nc, \Ec, \Cc)$, where
$\Nc=[1:N]$ is the set of nodes, $\Ec\subseteq [1:N]\times[1:N]$ is
the set of edges, and $\Cc=\{C_{jk} \in \Real^+:(j, k)\in \Ec\}$ is
the set of link capacities. Each edge $(j,k) \in \Ec$ carries an input
symbol $x_{jk} \in \Xc_{jk}$ with link capacity $C_{jk}=\log
|\Xc_{jk}|$, resulting in the channel output at node $k$ as
$Y_k=\{X_{jk}: (j, k)\in \Ec\}$. By setting $\Yh_k = Y_k$ for all $k$
and evaluating Theorem~\ref{thm:mmn} with the uniform pmf on $X^N$, it
can be easily shown that inner bound~\eqref{eq:mmn} coincides with the
cutset bound, and thus the capacity region is the set of rate tuples
$(R_1, \ldots, R_N)$ such that
\begin{equation}
R(\Sc) \leq \sum_{\substack{(j, k)\in \Ec \\ j\in \Sc, k\in
    \Sc^c}}C_{jk} \label{eq:noiseless-lb}.
\end{equation}
This recovers previous results in~\cite{Ahlswede--Cai--Li--Yeung2000} for the single-source case
and~\cite{Dana--Gowaikar--Palanki--Hassibi--Effros2006} for the
multiple-source case.

\medskip

\noindent {\em Relay channels:} Consider the relay channel
$p(y_2,y_3|x_1,x_2)$. It can be easily shown that the inner
bound~\eqref{eq:mmn} reduces to the alternative characterization of
the compress--forward lower bound in \eqref{eq:emz-cf}.

\medskip

\noindent {\em Erasure networks:} Consider the erasure multiple-source
multicast network in which the channel output at node $k \in [1:N]$ is
$Y_k=\{Y_{j k}: j\in [1:N]\}$, where $Y_{jk}=\varepsilon$ if it is
erased, and $Y_{jk}=X_j$, otherwise.  Assume further that the network
erasure pattern is known at the destination nodes. Taking $\Yh_k =
Y_k$, $k \in [1:N]$ and the uniform pmf on $X^N$ as in the noiseless
case, inner bound~\eqref{eq:mmn} reduces to
\begin{equation}
R(\Sc) \leq \sum_{j \in \Sc} \bigl(\log |\Xc_{j}|
(1-P\{\text{link $(j,k)$ is erased for all $k \in
  \Sc^c$}\})\bigr). \label{eq:erasure}
\end{equation}
It can be also shown that the inner bound coincides with the cutset
bound and thus characterizes the capacity region. This recovers the
previous result
in~\cite{Dana--Gowaikar--Palanki--Hassibi--Effros2006}.

\noindent {\em Deterministic networks:} Suppose $Y_k = g_{k}(X_1,
\ldots, X_N)$, $k \in [1:N]$.  By setting $\Yh_k = Y_k$, $k \in [1:N]$,
Theorem~\ref{thm:sd-mmn} implies that a rate tuple $(R_1, \ldots,
R_N)$ is achievable for the deterministic network if there exists some
pmf $p(q)\prod_{k=1}^N p(x_k|q)$ such that
\begin{equation}\label{eq:deterministic-lb}
R(\Sc) < I(X(\Sc); Y(\Sc^c)|X(\Sc^c), Q) 
= H(Y(\Sc^c)|X(\Sc^c), Q) 
\end{equation}
for all $\Sc \subseteq [1:N]$ with $\Sc^c \cap \Dc \neq \emptyset$.
This recovers previous results in~\cite{Avestimehr--Diggavi--Tse2009}
for the single-source case and in~\cite{Perron2009} for the
multiple-source case. Note that the lower
bound~\eqref{eq:deterministic-lb} is tight when the cutset bound is
attained by the product pmf, for example, as in the deterministic
network without interference~\cite{Ratnakar--Kramer2006} or the
finite-field linear deterministic network $Y_k = \sum_{j=1}^N g_{jk}
X_j$~\cite{Avestimehr--Diggavi--Tse2009}.

\medskip

Note that in all the above special cases, the channel output at node $k$ can be expressed as a deterministic function of the input symbols $(X_1,
\ldots, X_N)$ and the destination output symbol $Y_d$, i.e.,
\begin{equation}\label{eq:semidet}
Y_k = g_{dk}(X_1, \ldots, X_N, Y_d)\quad \text{for every } 
k\in [1:N] \text{ and } d\in \Dc.
\end{equation}
Under this structure, the inner bound in
Theorem \ref{thm:mmn} can be simplified by substituting $\Yh_k = Y_k$ for $k \in [1:N]$ in
\eqref{eq:mmn} to obtain the following generalization.
\begin{corollary} \label{thm:sd-mmn}
Let $\Dc = \Dc_1 = \cdots = \Dc_N$.  A rate tuple $(R_1, \ldots, R_N)$
is achievable for the semideterministic DMN~\eqref{eq:semidet} if
there exists some joint pmf $p(q) \prod_{k=1}^N p(x_k|q)$ such that
\begin{equation}
R(\Sc) <  I(X(\Sc); Y(\Sc^c)| X(\Sc^c), Q) \label{eq:sd-lb}
\end{equation}
for all $\Sc\subseteq [1:N]$ with $\Sc^c \cap \Dc \neq \emptyset$. 
\end{corollary}

We also show in Appendix~\ref{app:kgg} that our noisy network coding
scheme can strictly outperform the extension of the original
compress--forward scheme for the relay channel to networks
in~\cite[Th 3]{Kramer--Gastpar--Gupta2005}.

\medskip

\noindent{\bf General multiple-source networks:} We extend the noisy
network coding theorem to general multiple-source networks.  As a
first step, we note that Theorem~\ref{thm:mmn} continues to hold for
general networks with {\em multicast completion} of destination nodes,
that is, when every message is decoded by all destination nodes $\Dc =
\cup_{k=1}^N \Dc_k$.  Thus, we can obtain an inner bound on the
capacity region for the DMN in the same form as \eqref{eq:mmn} with
$\Dc=\cup_{k=1}^N \Dc_k$.

This multicast-completion inner bound can be improved by noting that
noisy network coding transforms a multi-hop relay network $p(y^N|x^N)$ into a
single-hop interference network $p(\yt^N|x^N)$, where the effective
channel output at decoder $k$ is $\Yt_k = (X_k, Y_k, \Yh_1,\ldots,
\Yh_N)$ and the compressed channel outputs $(\Yh_1, \ldots, \Yh_N)$
are conveyed to decoders with some rate penalty. This observation
leads to a modified decoding rule that does not require each
destination to decode unintended messages correctly, resulting in the following improved inner bound.

\begin{theorem} \label{thm:dmn-mac}
A rate tuple $(R_1, \ldots, R_N)$ is achievable for the DMN if there
exists some joint pmf $p(q)\prod_{k=1}^{N}p(x_k|q) p(\yh_k|y_k,x_k,q)$
such that
\begin{equation}\label{eq:dmn-mac}
R(\Sc) < \min_{d\in \Sc^c \cap \Dc (\Sc)} I(X(\Sc); \Yh (\Sc^c),
Y_d|X(\Sc^c), Q) - I(Y(\Sc);\Yh(\Sc)|X^{N}, \Yh(\Sc^c), Y_d, Q)
\end{equation}
for all cutsets $\Sc\subseteq [1:N]$ with $\Sc^c \cap \Dc(\Sc) \neq
\emptyset$, where $\Dc(\Sc):= \cup_{k\in \Sc} \Dc_k$.
\end{theorem}
The proof of this theorem is given in Subsection~\ref{subsec:dmn-mac}.

As an alternative, each destination node can simply treat interference
as noise rather than decoding it. Using this approach, we establish the 
following inner bound on the capacity region.
\begin{theorem} \label{thm:dmn-noise}
A rate tuple $(R_1, \ldots, R_N)$ is achievable for the DMN if there
exists some joint pmf $p(q)\prod_{k=1}^{N}p(u_k, x_k|q)
p(\yh_k|y_k,u_k,q)$ with
\begin{align}\label{eq:dmn-noise}
R(\Tc) < & I( X(\Tc), U(\Sc); \Yh (\Sc^c),
Y_{d}| X(\Tc^c), U(\Sc^c), Q) - I(Y(\Sc);\Yh(\Sc)| X(\Sc_d), U^{N}, \Yh(\Sc^c), Y_{d}, Q)
\end{align}
for all $\Sc\subseteq [1:N]$, $d\in \Dc(\Sc)$, and $\Sc\cap \Sc_d
\subseteq \Tc \subseteq \Sc_d$ such that $\Sc^c\cap \Dc(\Sc) \neq
\emptyset$, where $\Tc^c = \Sc_d \backslash \Tc$.
\end{theorem}

Unlike the coding schemes in Theorems~\ref{thm:mmn}
and~\ref{thm:dmn-mac} where each node maps both its own message and
the compression index to a single codeword, here each node applies
superposition coding~\cite{Cover1972} for forwarding the compression
index along with its own message. (Note that when a node does not have
its own message and it acts only as a relay, there is no difference in
the relay operation from the previous schemes.)  The details are given
in Subsection~\ref{subsec:dmn-noise}.

\medskip

\noindent{\bf Gaussian networks:} In Section~\ref{sec:gaussian}, we
present an extension of the above results to Gaussian networks and
compare the performance of noisy network coding to other specialized
coding schemes for two popular Gaussian networks.

Consider the Gaussian network 
\begin{equation} \label{eq:gaussian}
Y^N= G X^N+Z^N,
\end{equation}
where $G \in \Real^{N\times N}$ is the channel gain matrix and $Z^N$
is a vector of independent Gaussian random variables with zero mean
and unit variance. We further assume average power constraint $P$ on
each sender $X_k$.

In Subsection~\ref{subsec:theorem4}, we establish the following result
on the multicast capacity region of this general Gaussian network.

\begin{theorem} \label{thm:awgn-mmn}
Let $\Dc = \Dc_1 = \cdots = \Dc_N$. For any rate tuple
$(R_1,\ldots,R_N)$ in the cutset outer bound, there exists $(R_1',
\ldots, R_N')$ in the inner bound in Theorem~\ref{thm:mmn} for the 
AWGN network~\eqref{eq:gaussian}
such that
\[
\sum_{k\in \Sc} (R_k - R_k') \le 
\frac{|\Sc|}{2} + \frac{\min\{|\Sc|,|\Sc^c|\}}{2}\log(2|\Sc|)
\]
for all $\Sc\subseteq [1:N]$ with $\Sc^c \cap \Dc \neq \emptyset$.
\end{theorem}
This theorem implies that the gap between the cutset bound and our
inner bound is less than or equal to $(N/4) \log (2N)$ for $N > 3$,
regardless of the values of the channel gain matrix~$G$ and power
constraint~$P$.

\begin{figure}[h!]
\begin{center}
\small
\psfrag{a}[cb]{Sum rate}
\psfrag{b}[ct]{Relay location $d$}
\includegraphics[scale=0.46]{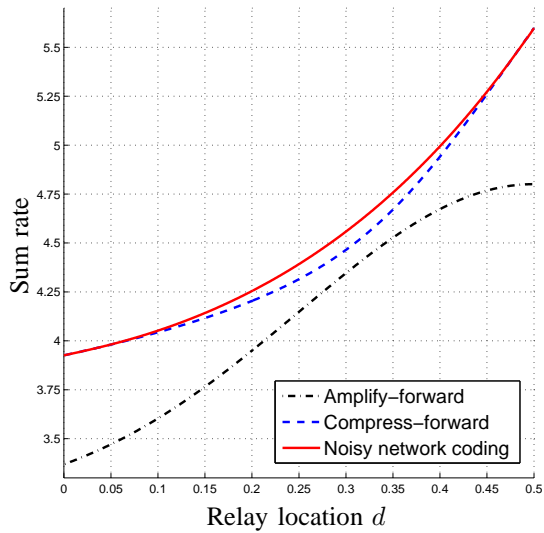}
\caption{Comparison of coding schemes for $P=10$.}
\label{fig:twrc}
\end{center}
\end{figure}

We also demonstrate through the following two examples that noisy
network coding can outperform previous coding schemes, some of which are developed specifically for
these channel models:

\noindent{\em Two-way relay channel (Subsection~\ref{subsec:awgn-twrc}):} Consider
the AWGN two-way relay channel
\begin{align}
Y_1&=g_{21}X_2+g_{31}X_3+Z_1, \nonumber\\
Y_2&=g_{12}X_1+g_{32}X_3+Z_2, \label{eq:awgn-twrc}\\
Y_3&=g_{13}X_1+g_{23}X_2+Z_3, \nonumber
\end{align}
where the channel gains are $g_{12} = g_{21} = 1$, $g_{13} = g_{31} =
d^{-\gamma/2}$ and $g_{23} = g_{32} = (1-d)^{-\gamma/2}$, and $d \in
[0,1]$ is the location of the relay node between nodes 1 and 2 (which
are unit distance apart).  Source nodes $1$ and $2$ wish to exchange
messages reliably with the help of relay node $3$. Various coding
schemes for this channel have been investigated
in~\cite{Rankov--Wittenben2006,
  Katti--Maric--Goldsmith--Katabi--Medard2007}. In
Figure~\ref{fig:twrc}, we compare the performance of noisy network
coding (Theorem 2) to amplify--forward (AF) and an extension of
compress--forward (CF) for $d \in [0,1/2]$ and $\gamma = 3$.  Note
that noisy network coding outperforms the other two schemes coinciding
with the compress--forward only when the relay is midway between nodes
1 and 2 ($d = 1/2$) and when it coincides with node 1 ($d=0$).

\vspace{-1em}

\begin{figure}[h!]
\begin{center}
\psfrag{Z1}[b]{$Z_4$}
\psfrag{Z3}[b]{$Z_3$}
\psfrag{w1}[b]{$\eta W_1$}
\psfrag{Z2}[t]{$Z_5$}
\psfrag{w2}[t]{$\eta W_2$}
\psfrag{X1}[r]{$X_1$}
\psfrag{X2}[r]{$X_2$}
\psfrag{Y1}[l]{$\,~Y_4$}
\psfrag{Y2}[l]{$\,~Y_5$}
\psfrag{t1}[l]{\hspace{-2pt}$Y_3$}
\psfrag{t2}[l]{$T_2$}
\psfrag{r0}[l]{$R_0$}
\psfrag{e}[b]{\small $g_{13}$}
\psfrag{f}[b]{\small $g_{23}$}
\psfrag{a}[rb]{\small $g_{24}$}
\psfrag{b}[rt]{\small $g_{15}$}
\psfrag{c}[t]{\small $g_{25}$}
\psfrag{d}[b]{\small $g_{14}$}
\includegraphics[width=0.4\linewidth]{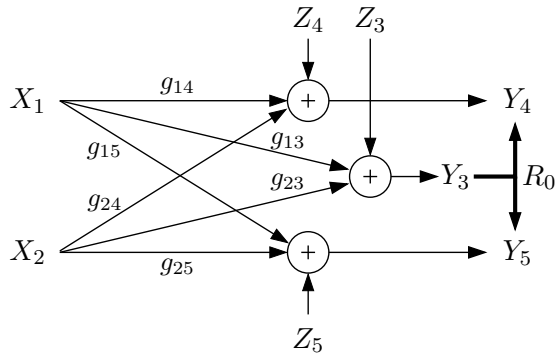}
\caption{AWGN interference relay channel.}
\label{fig:irc-prim}
\end{center}
\end{figure}

\noindent{\em Interference relay channel
  (Subsection~\ref{subsec:awgn-irc}):} Consider the AWGN interference
relay channel with orthogonal receiver components in
Figure~\ref{fig:irc-prim}.

\begin{figure}[h!]
\begin{center}
\small
\psfrag{a}[cb]{Sum rate}
\psfrag{b}[ct]{Power constraint $P$ (in dB)}
\includegraphics[scale=0.46]{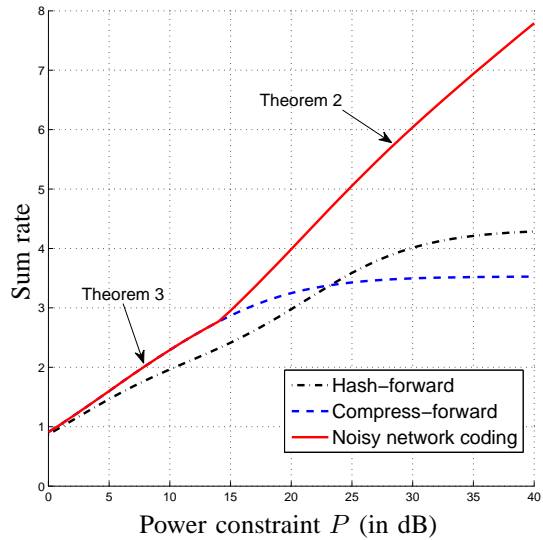}
\caption{Comparison of coding schemes for 
$g_{14}=g_{25}=1$, $g_{15}=g_{24}=g_{13}=0.5$, $g_{13}=0.1$.}
\label{fig:irc}
\end{center}
\end{figure}

The channel outputs are
\[
Y_j =g_{1j}X_1+g_{2j}X_2+Z_j,\quad j=3,4,5,
\]
where $g_{ij}$ is the channel gain of link $(i,j)$.  Source node 1
wishes to send a message to destination node 4, while source node 2
wishes to send a message to destination node 5.  Relay node 3 helps
the communication of this interference channel by sending some
information about $Y_3$ over a common noiseless link of rate $R_0$ to
both destination nodes. In Figure~\ref{fig:irc}, we compare noisy
network coding (Theorems 2 and 3) to compress--forward (CF) and
hash--forward (HF) in \cite{Razaghi--Yu2010}. The curve representing
noisy network coding depicts the maximum of achievable sum rates in
Theorems~\ref{thm:dmn-mac} and~\ref{thm:dmn-noise}. Note that,
although not shown in the figure, Theorem 3 alone outperforms the
other two schemes for all channel gains and power constraints. At high
signal-to-noise ratio (SNR), Theorem 2 provides further improvement,
since decoding other messages is a better strategy when interference
is strong.

\section{Noisy Network Coding for Multicast} \label{sec:mmn}

To illustrate the main idea of the noisy network coding scheme and
highlight the differences from the standard compress--forward coding
scheme~\cite{Cover--El-Gamal1979, Kramer--Gastpar--Gupta2005}, we
first prove Theorem~\ref{thm:mmn} for the 3-node relay channel and
then extend the proof to general multicast networks.
\medskip

Let $\xv_{kj}$ denote $(x_{k, (j-1)n+1}, \ldots, x_{k, jn })$,
$j\in[1:b]$; thus $x_{k}^{bn}= (x_{k1}, \ldots, x_{k, nb}) =
(\xv_{k1}, \ldots, \xv_{kb}) = \xv_k^b$. To send a message $m\in
[1:2^{nbR}]$, the source node transmits $\xv_{1j}(m)$ for each block
$j\in[1:b]$. In block $j$, the relay finds a ``compressed'' version
$\yvh_{2j}$ of the relay output $\yv_{2j}$ conditioned on $\xv_{2j}$,
and transmits a codeword $\xv_{2, j+1}(\yvh_{2j})$ in the next
block. After $b$ block transmissions, the decoder finds the correct
message $m\in [1:2^{nbR}]$ using $(\yv_{31}, \ldots, \yv_{3b})$ by
joint typical decoding for each of $b$ blocks simultaneously. The
details are as follows.

\medskip

\noindent\emph{Codebook generation:} Fix $p(x_1)p(x_2)p(\yh_2|y_2,x_2)$. We
randomly and independently generate a codebook for each block.

For each $j \in [1:b]$, randomly and independently generate $2^{nbR}$
sequences $\xv_{1j}(m),$ $m \in [1:2^{nbR}],$ each according to
$\prod_{i=1}^n p_{X_1}(x_{1,(j-1)n+i})$. Similarly, randomly and
independently generate $2^{n\Rh_2}$ sequences $\xv_{2j}(l_{j-1}),$
$l_{j-1} \in [1:2^{n\Rh_2}],$ each according to $\prod_{i=1}^n
p_{X_2}(x_{2, (j-1)n+i})$. For each $\xv_{2j}(l_{j-1}),$
$l_{j-1} \in [1:2^{n\Rh_2}],$ randomly and conditionally
independently generate $2^{n\Rh_2}$ sequences
$\yvh_{2j}(l_j|l_{j-1}),$ $l_j \in [1: 2^{n\Rh_2}],$ each
according to $\prod_{i=1}^n p_{\Yh_2 |X_2} (\yh_{2, (j-1)n+i}|x_{2,
  (j-1)n+i}(l_{j-1}))$.

This defines the codebook
\[
\Cc_j = \bigl\{\xv_{1j}(m), \xv_{2j}(l_{j-1}),
\yvh_{2j}(l_j|l_{j-1}): m \in [1:2^{nbR}], l_j, l_{j-1} \in
    [1:2^{n\Rh_2}]\bigr\}
\]
for $j \in [1:b]$.

Encoding and decoding are explained with the help of Table~\ref{tb:coding}.
\begin{table}[t]
\begin{center}
\small
\begin{tabular}{c|cccccc}
\hline
\textrm{Block} & 1 & 2 & 3 & $\cdots$ & $b-1$ & $b$\\
\hline%
$X_1$ &  $\xv_{11}(m)$ & $\xv_{12}(m)$ & $\xv_{13}(m)$ & $\ldots$ & $\xv_{1, b-1}(m)$ & $\xv_{1b}(m)$\\
$Y_{2}$ & $\yvh_{21}(l_1|1),l_1$ & $\yvh_{22}(l_2|l_1),l_2$ &
$\yvh_{23}(l_3|l_2),l_3$ & $\ldots$ & $\yvh_{2, b-1}(l_{b-1}|l_{b-2}),l_{b-1}$ &
$\yvh_{2b}(l_b|l_{b-1}),l_b$ \\
$X_2$ & $\xv_{21}(1)$ &  $\xv_{22}(l_1)$ & $\xv_{23}(l_2)$ &  $\ldots$ & 
$\xv_{2, b-1}(l_{b-2})$ & $\xv_{2b}(l_{b-1})$\\
$Y_3$ & $\emptyset$ & $\emptyset$ & $\emptyset$ & $\ldots$ & $\emptyset$ & $\mh$\\
\hline
\end{tabular}
\end{center}
\caption{Noisy network coding for the relay channel.} \label{tb:coding}
\end{table}
\medskip

\noindent\emph{Encoding:} Let $m$ be the message to be sent. The relay, upon
receiving $\yv_{2j}$ at the end of block $j\in[1:b]$, finds an
index $l_j$ such that 
\[
(\yvh_{2j}(l_j|l_{j-1}),\yv_{2j}, \xv_{2j}(l_{j-1})) \in
\aepvar, 
\]
where $l_{0}=1$ by convention. If there is more than one such index,
choose one of them at random. If there is no such index, choose an
arbitrary index at random from $[1:2^{n\Rh_2}]$. The codeword
pair $(\xv_{1j}(m), \xv_{2j}(l_{j-1}))$ is transmitted in block $j\in
[1:b]$.

\medskip
\noindent\emph{Decoding:} Let $\e > \e'$. At the end of block $b$, the
decoder finds a unique message $\mh\in [1:2^{nbR}]$ such that
\begin{align*}
(\xv_{1j}(\mh), \yvh_{2j}(\lh_j|\lh_{j-1}), \xv_{2j}(\lh_{j-1}), \yv_{3j}) &\in \aep
\quad\text{ for all } j \in [1:b]
\end{align*}
for some $\lh_1,\lh_2,\ldots,\lh_b$. If there is none or more than one
such message, it declares an error.

\medskip
\noindent\emph{Analysis of the probability of error:} Let $M$ denote the message sent at the source node and $L_j$ denote the indices chosen by the relay at block $j\in [1:b]$. Define
\begin{align*}
\Ec_0 &:= \bigcup_{j=1}^b\bigl\{ (\Yvh_{2j}(l_j|L_{j-1}), \Xv_{2j}(L_{j-1}), \Yv_{2j})\not \in \aepvar \text{ for all } l_j\bigr\},\\
\Ec_m &:=\bigl\{(\Xv_{1j}(m), \Yvh_{2j}(l_j|l_{j-1}), \Xv_{2j}(l_{j-1}), \Yv_{3j}) \in \aep,\, j\in[1:b] 
\text{ for some } l_1,l_2,\ldots,l_b\bigr\}.
\end{align*}
To bound the probability of error, assume without loss of generality
that $M=1$.  Then the probability of error is upper bounded by
\begin{align*}
\P(\Ec)
\leq \P(\Ec_0)+\P(\Ec_0^c \cap \Ec_1^c)+\P(\cup_{m\neq 1} \Ec_m).
\end{align*}
By the covering lemma~\cite{El-Gamal--Kim2010}, $\P(\Ec_0)\to 0$ as $n
\to \infty$, if $R_2 > I(\Yh_2;Y_2|X_2)+\delta(\e')$. By the conditional
typicality lemma~\cite{El-Gamal--Kim2010}, $\P(\Ec_0^c\cap \Ec_1^c
)\to 0$ as $n\to \infty$.

To bound $\P(\cup_{m\neq 1} \Ec_m)$, assume without loss of generality
that $(L_1,\ldots,L_b) = (1,\ldots,1)$; recall the symmetry of the
codebook construction. 
For $j \in [1:b]$, $m \in [1:2^{nbR}]$, and $l_{j-1}, l_j
\in [1:2^{n\Rh_2}]$, define the events
\[
\Ac_{j}(m, l_{j-1}, l_j) := 
\bigl\{(\Xv_{1j}(m), \Yvh_{2j}(l_j|l_{j-1}), 
\Xv_{2j}(l_{j-1}), \Yv_{3j}) \in \aep\bigr\}.
\]
Then, 
\begin{align}
\P(\Ec_m)&=\P( \cup_{l^b} \cap_{j=1}^b \Ac_j(m, l_{j-1}, l_j))\nonumber\\
&\le 
\sum_{l^b}
\P( \cap_{j=1}^b \Ac_j(m, l_{j-1}, l_j)) \nonumber\\
&= 
\sum_{l^b} 
\prod_{j=1}^b \P( \Ac_j(m, l_{j-1}, l_j)) \label{eq:codeInd1}\\
&\le  
\sum_{l^b}
\prod_{j=2}^b \P( \Ac_j(m, l_{j-1}, l_j)), \nonumber
\end{align}
where equality~\eqref{eq:codeInd1} follows since the codebook is
generated independently for each block $j$ and the channel is
memoryless.  Note that if $m \ne 1$ and $l_{j-1} = 1$, then
$\Xv_{1j}(m)\sim \prod_{i=1}^n p_{X_1}(x_{1, (j-1)n+i})$ is
independent of $(\Yvh_{2j}(l_j|l_{j-1}), \Xv_{2j}(l_{j-1}), \Yv_{3j})$
(given $L_{j-1}=L_j=1$). Hence, by the joint typicality
lemma~\cite{El-Gamal--Kim2010},
\begin{align}
\P (\Ac_j(m, l_{j-1}, l_j)) 
&= \P \bigl\{(\Xv_{1j}(m), \Yvh_{2j}(l_j|l_{j-1}), 
\Xv_{2j}(l_{j-1}), \Yv_{3j}) \in \aep\bigr\} \nonumber \\
&\le 2^{-n (I(X_1; \Yh_2, Y_3 | X_2) - \delta(\e))} \nonumber\\
&=: 2^{-n (I_1 - \delta(\e))}. \label{eq:bnd1}
\end{align}
Similarly, if $m \ne 1$ and $l_{j-1} \ne 1$, then 
\[
(\Xv_{1j}(m), \Xv_{2j}(l_{j-1}), \Yvh_{2j}(l_{j}| l_{j-1})) \sim \prod_{i=1}^n p_{X_1}(x_{1, (j-1)n+i}) p_{X_2, \Yh_2}(x_{2, (j-1)n+i}, \yh_{2, (j-1)n+i})
\] 
is independent of $\Yv_{3j}$ (given $L_{j-1}= L_j=1$). 
Hence, by the joint typicality lemma
\begin{equation}
\P (\Ac_j(m, l_{j-1}, l_j) ) \le 2^{-n (I(X_1,X_2;Y_3) 
+ I(\Yh_2;X_1,Y_3|X_2)- \delta(\e))}
=: 2^{-n (I_2- \delta(\e))}.  \label{eq:bnd2}
\end{equation}
If the binary sequence $l^{b-1}$ has $k$ 1s, then by \eqref{eq:bnd1}
and \eqref{eq:bnd2},
\begin{align*}
\prod_{j=2}^b \P( \Ac_j(m, l_{j-1}, l_j))\le 
2^{ -n \left(k I_1
+ (b-1-k) I_2- (b-1) \delta(\e)\right)}.
\end{align*}
Therefore
\begin{align*}
\sum_{l^b}
\prod_{j=2}^b \P( \Ac_j(m, l_{j-1}, l_j))
&= \sum_{l_b}\sum_{l^{b-1}}
\prod_{j=2}^b \P( \Ac_j(m, l_{j-1}, l_j))\\
&\le 
 \sum_{l_b} 
\sum_{k=0}^{b-1} \mbinom{b-1}{k} 2^{n(b-1-k) \Rh_2}
2^{ -n \left(k I_1
+ (b-1-k) I_2- (b-1) \delta(\e)\right)} \\
&= 
 \sum_{l_b} 
\sum_{k=0}^{b-1}\mbinom{b-1}{k}
2^{ -n \left(k I_1
+ (b-1-k) (I_2 - \Rh_2) - (b-1) \delta(\e)\right)} \\
&\le 
\sum_{l_b} 
\sum_{k=0}^{b-1}\mbinom{b-1}{k}
2^{ -n\left((b-1) 
(\min \{ I_1,\,
I_2- \Rh_2\} - \delta(\e))\right)} \\
&=
2^{n\Rh_2}2^{b-1}\cdot 2^{-n\left((b-1)(
\min \{ I_1,\,
I_2- \Rh_2\} - \delta(\e))\right)}.
\end{align*}
Thus, 
\[
\sum_{m\ne 1} \sum_{l^b}
\prod_{j=2}^b \P( \Ac_j(m, l_{j-1}, l_j) \mid L_{j-1}= L_j=1) \to 0
\]
 as $n \to \infty$, provided that
\[
\hspace{-5pt} R < \frac{b-1}{b} 
(\min \{ I_1,\,
I_2- \Rh_2\} - \delta(\e)) - \frac{1}{b} \Rh_2.
\]
Finally, by eliminating $\Rh_2 > I(\Yh_2;Y_2|X_2)+\delta(\e')$ and
letting $b \to \infty$, we have shown the achievability of any rate
\[
R < \min \{I(X_1; \Yh_2,Y_3|X_2),\,
I(X_1,X_2;Y_3) - I(\Yh_2;Y_2|X_1,X_2,Y_3)\} - \delta(\e) - \delta(\e').
\]
This concludes the proof of Theorem 1 for the special case of the
relay channel.

\medskip

We now describe the noisy network coding scheme for multiple-source
multicast over a general DMN $p(y^N|x^N)$. For simplicity of notation,
we consider the case $Q = \emptyset$. Achievability for an arbitrary
time-sharing random variable $Q$ can be proved using the coded time
sharing technique~\cite{El-Gamal--Kim2010}.

\medskip
 
\noindent\emph{Codebook generation:} Fix $\prod_{k=1}^N
p(x_k)p(\yh_k|y_k,x_k)$. We randomly and independently generate a
codebook for each block.
For each $j\in [1:b]$ and $k\in [1:N]$, randomly and independently
generate $2^{nbR_k}\times2^{n\Rh_k}$ sequences $\xv_{k, j}(m_k, l_{k,
  j-1}),$ $m_k \in [1:2^{nbR_k}]$, $l_{k, j-1}\in [1:2^{n\Rh_k}]$,
each according to $\prod_{i=1}^n p_{X_k}(x_{k, (j-1)n+i})$. For each
node $k \in [1:N]$ and each $\xv_{kj}(m_k, l_{k, j-1}),$ $m_k\in
[1:2^{nbR_k}],$ $l_{k, j-1} \in [1:2^{n\Rh_k}],$ randomly and
conditionally independently generate $2^{n\Rh_k}$ sequences
$\yvh_{kj}(l_{kj}|m_k, l_{k, j-1}),$ $l_{kj} \in [1: 2^{n\Rh_k}],$
each according to $\prod_{i=1}^n p_{\Yh_k |X_k} (\yh_{k,
  (j-1)n+i}|x_{k, (j-1)n+i}(m_k, l_{k, j-1}))$.  This defines the
codebook
\[
\Cc_j = \bigl\{\xv_{kj}(m_k, l_{k, j-1}),
\yvh_{kj}(l_{kj}|m_k, l_{k, j-1}): m_k \in [1:2^{nbR_k}], l_{kj}, l_{k,j-1} \in
    [1:2^{n\Rh_k}], k\in [1:N]\bigr\}
\]
for $j \in [1:b]$.

\medskip

\noindent\emph{Encoding:} Let $(m_1, \ldots, m_N)$ be the messages to be sent. Each node $k\in [1:N]$, upon receiving $\yv_{kj}$ at the end of block $j\in [1:b]$,
finds an index $l_{kj}$ such that
\[
(\yvh_{kj}(l_{kj}|m_k, l_{k,j-1}),\yv_{kj},\xv_{kj}(m_k, l_{k,j-1})) \in \aepvar,
\]
where $l_{k0}=1$, $k\in[1:N]$, by convention. If there is more than
one such index, choose one of them at random. If there is no such
index, choose an arbitrary index at random from $[1:2^{n\Rh_k}]$. Then
each node $k\in [1:N]$ transmits the codeword $\xv_{kj}(m_k, l_{k,
  j-1})$ in block $j \in [1:b]$.

\medskip

\noindent\emph{Decoding:} Let $\e > \e'$. At the end of block $b$,
decoder $d\in \Dc$ finds a unique index tuple $(\mh_{1d}, \ldots,
\mh_{Nd})$, where $\mh_{kd}\in[1:2^{nbR_k}]$ for $k\neq d$ and
$\mh_{dd}=m_d$, such that there exist some $(\lh_{1j}, \ldots,
\lh_{Nj})$, $\lh_{kj}\in[1:2^{n\Rh_k}]$, $k\neq d$ and
$\lh_{dj}=l_{dj}$, $j\in[1:b]$, satisfying
\begin{align*}
(\xv_{1j}(\mh_{1d}, \lh_{1,j-1}), \ldots,&\xv_{Nj}(\mh_{Nd}, \lh_{N,j-1}),\\
& \yvh_{1j}(\lh_{1j}| \mh_{1d}, \lh_{1,j-1}), \ldots, \yvh_{Nj}(\lh_{Nj}| \mh_{Nd}, \lh_{N,j-1}), \yv_{dj})\in\aep
\end{align*}
for all $j\in[1:b]$.

\medskip

\noindent\emph{Analysis of the probability of error:} Let $M_k$ denote
the message sent at node $k\in[1:N]$ and $L_{kj}$, $k\in [1:N]$, $j\in
[1:b]$, denote the index chosen by node $k$ for block $j$. To bound
the probability of error for decoder $d\in \Dc$, assume without loss
of generality that $(M_1,\ldots,M_N) =(1, \ldots, 1)=:\mathbf
1$. Define
\begin{align*}
\Ec_0 &:= \bigcup_{j=1}^b\bigcup_{k=1}^N
\bigl\{(\Yvh_{kj}(l_{kj}| 1, L_{k, j-1}), \Xv_{kj}(1, L_{k, j-1}), \Yv_{kj}) \not \in \aepvar \text{ for all } l_{kj}\bigr\}\\
\Ec_{\mv} &:=\bigl\{(\Xv_{1j}(m_{1}, l_{1,j-1}), \ldots,\Xv_{Nj}(m_{N}, l_{N,j-1}), \\
&\qquad\qquad\qquad\Yvh_{1j}(l_{1j}| m_{1}, l_{1,j-1}), \ldots, \Yvh_{Nj}(l_{Nj}| m_{N}, l_{N,j-1}), \Yv_{dj}) \in \aep, \\
&\qquad\qquad \qquad \qquad j\in [1:b],\text{ for some } (\lv_{1}, \ldots, \lv_{b}), \text{ where } l_{dj}=L_{dj}, j\in[1:b]\bigr\}.
\end{align*}
Here, $\lv_j = (l_{1j}, \ldots, l_{Nj})$ for $j\in[1:b]$.
Then the probability of error is upper bounded as
\begin{equation}
\P(\Ec) \leq \P(\Ec_0)+\P(\Ec_0^c \cap \Ec_{\mathbf 1}^c)+\P(\cup_{\mv \neq \mathbf 1} \Ec_{\mv}), \label{eq:three-error}
\end{equation}
where $\mv := (m_1, \ldots, m_N)$ such that $m_d = 1$. As in the
$3$-node case, by the covering lemma, $\P(\Ec_0)\to 0$ as $n \to
\infty$, if $\Rh_k > I(\Yh_k;Y_k|X_k)+\delta(\e')$, $k\in [1:N]$, and by
the conditional typicality lemma $\P(\Ec_0^c \cap \Ec_{\mathbf 1}^c
)\to 0$ as $n\to \infty$.  For the third term, assume without loss of
generality that $\Lv_1=\cdots = \Lv_b=\mathbf 1$, where
$\Lv_{j}:=(L_{1j}, \ldots, L_{Nj})$. Define the events
\begin{align*}
\Ac_{j}(\mv, \lv_{j-1}, \lv_j)
&:= 
\bigl\{(\Xv_{1j}(m_{1}, l_{1, j-1})
,\ldots, \Xv_{Nj}(m_{N}, l_{N, j-1}),\\
&\qquad  \Yvh_{1j}(l_{1j}|m_1, l_{1,j-1}), \ldots,
\Yvh_{Nj}(l_{Nj}|m_N, l_{N,j-1}),
\Yv_{dj}) \in \aep \bigr\}
\end{align*}
for $\mv\neq \mathbf 1$ and all $\lv_j$. 
Then, 
\begin{align}
	\P(\Ec_{\mv})&=\P( \cup_{\lv^b} \cap_{j=1}^b \Ac_j(\mv, \lv_{j-1}, \lv_j))\nonumber\\
&\le \sum_{\lv^b}
\P( \cap_{j=1}^b \Ac_j(\mv, \lv_{j-1}, \lv_j)) \nonumber\\
&= \sum_{\lv^b}
\prod_{j=1}^b \P( \Ac_j(\mv, \lv_{j-1}, \lv_j) ) \label{eq:codeInd2}\\
&\le  \sum_{\lv^b}
\prod_{j=2}^b \P( \Ac_j(\mv, \lv_{j-1}, \lv_j)), \nonumber
\end{align}
where \eqref{eq:codeInd2} follows since the codebook is generated
independently for each block $j$ and the channel is memoryless.

For each $\lv^b$ and $j \in [2:b]$, let $\Sc_j(\mv, \lv^b) \subseteq
[1:N]$ such that $\Sc_j(\mv, \lv^b)=\{k: m_k \neq 1 \text{ or }
l_{k,j-1}\neq 1\}$. Note that $\Sc_j(\mv, \lv^b)$ depends only on
$(\mv, \lv_{j-1})$ and hence we write it as $\Sc_j(\mv,
\lv_{j-1})$. We further define $\Tc(\mv) \subseteq [1:N]$ such that
$\Tc(\mv)=\{k: m_k \neq 1\}$. From the definitions we can see that
$\Tc(\mv) \subseteq \Sc_j(\mv, \lv_{j-1})$ and $d\in \Sc^c_j(\mv,
\lv_{j-1})\subseteq \Tc^c(\mv)$.

Define $\Xv_j(\Sc_j(\mv, \lv_{j-1}))$ to be the set of $\Xv_{kj}(m_k,
l_{k, j-1})$, $k\in \Sc_j(\mv, \lv_{j-1})$, where $m_k$ and $l_{k,
  j-1}$ are the corresponding elements in $\mv$ and $\lv^b$,
respectively. Similarly define $\Yvh_j(\Sc_j(\mv, \lv_{j-1}))$ and
$\Yv_j(\Sc_j(\mv, \lv_{j-1}))$.  Then, by the joint typicality lemma
and the fact that
\[
 \bigl(\Xv(\Sc_j(\mv, \lv_{j-1})), \Yvh(\Sc_j(\mv, \lv_{j-1}))\bigr) \sim \prod_{k\in \Sc_j(\mv, \lv_{j-1})} \prod_{i=1}^n p_{X_k}(x_{k, (j-1)n+i})\,  p_{\Yh_k|X_k}(\yh_{k, (j-1)n+i}|x_{k, (j-1)n+i})
\]  
is independent of $\bigl(\Xv(\Sc_j^c(\mv, \lv_{j-1})), \Yvh(\Sc_j^c(\mv, \lv_{j-1})), \Yv_{dj}\bigr)$ (given $\Lv_{j-1}=\Lv_j=\mathbf 1$), we have
\begin{equation}
\P (\Ac_j(\mv, \lv_{j-1}, \lv_j)) \leq 2^{-n (I_1(\Sc(\mv, \lv_{j-1})) + I_2(\Sc(\mv, \lv_{j-1})) - \delta(\e))}, \nonumber
\end{equation}
where 
\begin{align*}
&I_1(\Sc) :=I(X(\Sc); \Yh(\Sc^c), Y_d | X(\Sc^c)),\\
&I_2(\Sc):=\sum_{k \in \Sc} I(\Yh_k; \Yh(\Sc^c\cup\{k'\in\Sc: k' < k\}),Y_d, X^N|X_k).
\end{align*}

Furthermore, from the definitions of $\Tc(\mv)$ and $\Sc_j(\mv,
\lv_{j-1})$, if $\mv \neq \mathbf 1$ with $m_d=1$, then
\begin{align*}
\sum_{\lv_{j-1}}&\,
2^{-n (I_1(\Sc_j(\mv, \lv_{j-1}))  + I_2(\Sc_j(\mv, \lv_{j-1}))-\delta(\e))}\\
&\le
\sum_{\substack{\Sc \subseteq [1:N]\\ \Tc(\mv) \subseteq \Sc, d \in \Sc^c}}\sum_{\lv_{j-1}: \Sc_j(\mv, \lv_{j-1})= \Sc}  
2^{-n (I_1(\Sc_j(\mv, \lv_{j-1}))  + I_2(\Sc_j(\mv, \lv_{j-1})) - \delta(\e))} \\
&\le
\sum_{\substack{\Sc \subseteq [1:N]\\ \Tc(\mv) \subseteq \Sc, d \in \Sc^c}}
2^{-n (I_1(\Sc)  + I_2(\Sc)
-\sum_{k \in \Sc} \Rh_k - \delta(\e))} \\
&\le
2^{N-1}2^{-n (\min_{\Sc}(I_1(\Sc)  + I_2(\Sc)
-\sum_{k \in \Sc} \Rh_k - \delta(\e)))}, 
\end{align*}
where the minimum is over $\Sc\subseteq [1:N]$ such that $\Tc(\mv)\subseteq \Sc$ and $d\in \Sc^c$. Hence,
\begin{align}
\sum_{\mv \ne \mathbf 1}& \sum_{\lv^b} 
\prod_{j=2}^b \P( \Ac_j(\mv, \lv_{j-1}, \lv_j))\nonumber\\
&=
\sum_{\mv \ne \mathbf 1}\sum_{\lv_b}\sum_{\lv^{b-1}} 
\prod_{j=2}^b \P( \Ac_j(\mv, \lv_{j-1}, \lv_j))\nonumber\\
&=
\sum_{\mv \ne \mathbf 1}\sum_{\lv_b} 
\prod_{j=2}^b \sum_{\lv_{j-1}} \P( \Ac_j(\mv, \lv_{j-1}, \lv_j))\nonumber\\
&\le
 \sum_{\mv\ne \mathbf 1}\sum_{\lv_b}
\prod_{j=2}^{b} \left( \sum_{\lv_{j-1}}
2^{-n (I_1(\Sc_j(\mv, \lv_{j-1}))  + I_2(\Sc_j(\mv, \lv_{j-1}))-\delta(\e))}
\right) \nonumber\\
&\le
\sum_{\substack{\Tc \subseteq [1:N] \\ \Tc \neq \emptyset, d\in \Tc^c}}2^{\sum_{k\in \Tc}nbR_k}\,2^{\sum_{k\neq d} n\Rh_k}2^{(N-1)(b-1)}\, 2^{n\left(- (b-1) \min_{\Sc} (I_1(\Sc) + I_2(\Sc) -
\sum_{k \in \Sc} \Rh_k - \delta(\e))\right)}, \label{eq:errorbnd}
\end{align} 
where the minimum in \eqref{eq:errorbnd} is over $\Sc \subseteq [1:N]$ such that $\Tc \subseteq \Sc, d \in \Sc^c$.
Hence, \eqref{eq:errorbnd} tends to zero as $n \to \infty$ if
\[
R(\Tc) < \frac{b-1}{b} \left(\min_{\substack{\Sc \subseteq [1:N]\\ \Tc \subseteq \Sc, d \in \Sc^c}} \left(I_1(\Sc) + I_2(\Sc) -
\sum_{k\in\Sc} \Rh_k \right)- \delta(\e)\right) - \frac{1}{b} \left(\sum_{k\neq d}
\Rh_k\right)
\]
for all $\Tc \subseteq [1:N]$ such that $\Tc\neq \emptyset$ and $d\in \Tc^c$.
By eliminating $\Rh_k > I(\Yh_k;Y_k|X_k)+\delta(\e')$ and
letting $b \to \infty$, the probability of error tends to zero as $n \to \infty$ if
\[
R(\Tc) < \min_{\substack{\Sc \subseteq [1:N]\\ \Tc \subseteq \Sc, d \in \Sc^c}} \left(I_1(\Sc) + I_2(\Sc) - \sum_{k \in \Sc} I(\Yh_k;Y_k|X_k)
\right)
- (N-1)\delta(\e') - \delta(\e)
\]
for all $\Tc \subseteq [1:N]$ such that $d\in \Tc^c$.
Finally, note that
\begin{align*}
I_2(\Sc) - \sum_{k \in \Sc} I(\Yh_k;Y_k|X_k)
&= 
-\sum_{k \in \Sc}
I(\Yh_k; Y_k | X^N, \Yh(\Sc^c), Y_d, \Yh(\{k'\in\Sc:k' < k\}))\\
&= 
-\sum_{k \in \Sc}
I(\Yh_k; Y(\Sc) | X^N, \Yh(\Sc^c), Y_d, \Yh(\{k'\in\Sc:k' < k\}))\\
&=
-I(\Yh(\Sc); Y(\Sc) | X^N, \Yh(\Sc^c), Y_d).
\end{align*}
Therefore, the probability of error tends to zero as $n \to \infty$ if
\begin{align}
R(\Tc) < I(X(\Sc);\Yh(\Sc^c), & Y_d|X(\Sc^c)) - I(\Yh(\Sc); Y(\Sc) | X^N, \Yh(\Sc^c), Y_d) - (N-1)\delta(\e') -\delta(\e) \label{eq:condition1}
\end{align}
for all $\Sc, \Tc \subseteq [1:N]$ such that $\emptyset \neq \Tc
\subseteq \Sc$ and $d\in \Sc^c$. Since for every $\Sc\subseteq [1:N]$
such that $\Sc\neq \emptyset$ and $d\in \Sc^c$ the inequalities with
$\Tc \subsetneq \Sc$ are inactive due to the inequality with $\Tc
=\Sc$ in \eqref{eq:condition1}, the set of inequalities can be further
simplified to
\begin{equation}
R(\Sc) <  I(X(\Sc);\Yh(\Sc^c), Y_d|X(\Sc^c)) - I(\Yh(\Sc); Y(\Sc) | X^N, \Yh(\Sc^c), Y_d) - (N-1)\delta(\e') -\delta(\e) \label{eq:condition2}
\end{equation}
for all $\Sc\subseteq [1:N]$ such that $d\in \Sc^c$. Thus, the
probability of decoding error tends to zero for each destination node
$d\in \Dc$ as $n\to \infty$, provided that the rate tuple satisfies
\eqref{eq:condition2}. 

Hence, by the union of events bound, the probability of error for {\em
  all} destinations tends to zero as $n \to \infty$ if the rate tuple
$(R_1, \ldots, R_N)$ satisfies
\begin{align*}
R(\Sc)< \min_{d\in \Sc^c\cap \Dc} I(X(\Sc); \Yh (\Sc^c), Y_d|X(\Sc^c)) - I(Y(\Sc);\Yh(\Sc)|&X^{N}, \Yh(\Sc^c), Y_d)
\end{align*}
for all $\Sc\subseteq [1:N]$ such that $\Sc^c\cap \Dc \neq \emptyset$
for some $\prod_{k=1}^{N} p(x_k)p(\yh_k|y_k,x_k)$. Finally, by coded
time sharing, the probability of error tends to zero as $n \to \infty$
if the rate tuple $(R_1, \ldots, R_N)$ satisfies
\begin{align*}
R(\Sc)< \min_{d\in \Sc^c\cap \Dc} I(X(\Sc); \Yh (\Sc^c),
Y_d|X(\Sc^c),Q) - I(Y(\Sc);\Yh(\Sc)|&X^{N}, \Yh(\Sc^c), Y_d,Q)
\end{align*}
for all $\Sc\subseteq [1:N]$ such that $\Sc^c\cap \Dc \neq \emptyset$
for some $\prod_{k=1}^{N} p(q)p(x_k|q)p(\yh_k|y_k,x_k,q)$.  This
completes the proof of Theorem~\ref{thm:mmn}.


\section{Extensions to General Multiple-Source Networks}\label{sec:general}

\subsection{Proof of Theorem 2 via Multicast Completion with Implicit Decoding}
\label{subsec:dmn-mac}

We modify the decoding rule in the previous section to establish
Theorem~\ref{thm:dmn-mac} as follows.

\medskip
\noindent\emph{Decoding:} At the end of block $b$, decoder $d\in
\cup_{k=1}^N\Dc_k$ finds a unique index tuple $\{\mh_{kd}: k\in
\Sc_d\}$ such that there exist some $(\mh_{kd}: k \in \Sc_d^c)$ and
$(\lh_{1j}, \ldots, \lh_{Nj})$, where $\mh_{kd}\in[1:2^{nbR_k}]$ for
$k\neq d$, $\mh_{dd}=m_d$, $\lh_{kj}\in[1:2^{n\Rh_k}]$ for $k\neq d$,
and $\lh_{dj}=l_{dj}$, $j\in[1:b]$, satisfying
\begin{align*}
(\xv_{1j}(\mh_{1d}, \lh_{1,j-1}), \ldots,&\xv_{Nj}(\mh_{Nd}, \lh_{N,j-1}),\\
& \yvh_{1j}(\lh_{1j}| \mh_{1d}, \lh_{1,j-1}), \ldots, \yvh_{Nj}(\lh_{Nj}| \mh_{Nd}, \lh_{N,j-1}), \yv_{dj})\in\aep
\end{align*}
for all $j\in[1:b]$.

The analysis of the probability of error is similar to that for
Theorem~\ref{thm:mmn} in Section~\ref{sec:mmn}. For completeness, the
details are given in Appendix~\ref{app:theorem2}.

\subsection{Proof of Theorem 3 via Treating Interference as Noise}
\label{subsec:dmn-noise}

\noindent\emph{Codebook generation:} Again we consider the case $Q =
\emptyset$.  Fix $\prod_{k=1}^N p(u_k, x_k)p(\yh_k|y_k,u_k)$. We
randomly and independently generate a codebook for each block.  For
each $j\in [1:b]$ and $k\in [1:N]$, randomly and independently
generate $2^{n\Rh_k}$ sequences $\uv_{kj}(l_{k, j-1})$, $l_{k, j-1}\in
[1:2^{n\Rh_k}]$, each according to $\prod_{i=1}^n p_{U_k}(u_{k,
  (j-1)n+i})$. For each $k \in [1:N]$ and each $\uv_{kj}(l_{k, j-1})$,
$l_{k, j-1} \in [1:2^{n\Rh_k}],$ randomly and conditionally
independently generate $2^{nbR_k}$ sequences $\xv_{kj}(m_k| l_{k,
  j-1})$, $m_k \in [1:2^{nbR_k}]$, each according to $\prod_{i=1}^n
p_{X_k |U_k} (x_{k, (j-1)n+i}|u_{k, (j-1)n+i}(l_{k, j-1}))$.  For each
$k \in [1:N]$ and each $\uv_{kj}(l_{k, j-1}),$ $l_{k, j-1} \in
[1:2^{n\Rh_k}],$ randomly and conditionally independently generate
$2^{n\Rh_k}$ sequences $\yvh_{kj}(l_{kj}| l_{k, j-1}),$ $l_{kj} \in
[1: 2^{n\Rh_k}],$ each according to
$
\prod_{i=1}^n p_{\Yh_k |U_k} (\yh_{k, (j-1)n+i}|u_{k, (j-1)n+i}(l_{k,
  j-1}))$.
This defines the codebook
\[
\Cc_j \!= \!\bigl\{ \uv_{kj}(l_{k,j-1}), \xv_{kj}(m_k| l_{k, j-1}),
\yvh_{kj}(l_{kj}|l_{k, j-1}): m_k \in [1:2^{nbR_k}], l_{kj}, l_{k,j-1} \in
    [1:2^{n\Rh_k}], k\in [1:N]\bigr\}
\]
for $j \in [1:b]$.

\medskip
\noindent\emph{Encoding:} Let $(m_1, \ldots, m_N)$ be the messages to be
sent. Each node $k\in [1:N]$, upon receiving $\yv_{kj}$ at the end of
block $j\in [1:b]$, finds an index $l_{kj}$ such that
\[
(\yvh_{kj}(l_{kj}|l_{k,j-1}),\yv_{kj},\uv_{kj}(l_{k,j-1})) \in \aepvar,
\]
where $l_{k0}=1$, $k\in[1:N]$, by convention. If there is more than
one such index, choose one of them at random. If there is no such
index, choose an arbitrary index at random from $[1:2^{n\Rh_k}]$. Then
each node $k\in [1:N]$ transmits the codeword $\xv_{kj}(m_k| l_{k,
  j-1})$ in block $j \in [1:b]$.

\medskip

Similarly as before, decoding is done by simultaneous joint typical
decoding, however, since we are treating interference as noise,
codewords corresponding to the unintended messages $(m_{k}: k \in
\Sc_d^c)$ are discarded, which leads to the following.

\medskip
\noindent\emph{Decoding:} At the end of block $b$, decoder $d\in
\cup_{k=1}^N\Dc_k$ finds a unique index tuple $(\mh_{kd}: k\in \Sc_d)$
such that there exist some $(\lh_{1j}, \ldots, \lh_{Nj})$, where
$\mh_{kd}\in[1:2^{nbR_k}]$ and $k\neq d$ and $\mh_{dd}=m_d$,
$\lh_{kj}\in[1:2^{n\Rh_k}]$, $k\neq d$ and $\lh_{dj}=l_{dj}$,
$j\in[1:b]$, satisfying
\begin{align*}
&((\xv_{kj}(\mh_{kd}| \lh_{k,j-1}): k\in \Sc_d), \uv_{1j}(\lh_{1,j-1}), \ldots,\uv_{Nj}(\lh_{N,j-1}),\\
&\qquad\qquad\qquad\qquad
\yvh_{1j}(\lh_{1j}|\lh_{1,j-1}), \ldots, \yvh_{Nj}(\lh_{Nj}| \lh_{N,j-1}), \yv_{dj})\in\aep
\end{align*}
for all $j\in[1:b]$.

The analysis of the probability of error is delegated to
Appendix~\ref{app:theorem3}.


\section{Gaussian Networks}\label{sec:gaussian}
We consider the additive white Gaussian noise (AWGN) network in which
the channel output vector for an input vector $X^N$ is $Y^N= G
X^N+Z^N$, where $G \in \Real^{N\times N}$ is the channel gain matrix
and $Z^N$ is a vector of independent additive white Gaussian noise
with zero mean and unit variance. We assume average power constraint
$P$ on each sender, i.e.,
\[
\sum_{i=1}^n \E \left(x^2_{ki}(m_k, Y^{i-1}_k)\right) \leq nP
\]
for all $k\in [1:N]$ and $m_k\in [1:2^{nR_k}]$. For each cutset
$\Sc\subseteq [1:N]$, define a channel gain submatrix $G(\Sc)$ such
that
\[
Y(\Sc^c)= G(\Sc) X(\Sc) + G'(\Sc) X(\Sc^c) + Z(\Sc^c).
\]
In the following subsection, we prove Theorem~\ref{thm:awgn-mmn}. In
Subsections~\ref{subsec:awgn-twrc} and~\ref{subsec:awgn-irc}, we provide the capacity
inner bounds for the AWGN two-way relay channel and the AWGN
interference relay channel used in Figures~\ref{fig:twrc}
and~\ref{fig:irc}.


\subsection{AWGN Multicast Capacity Gap (Proof of Theorem 4)}
\label{subsec:theorem4}

The cutset outer bound for the AWGN multiple-source multicast network
simplifies to the set of rate tuples such that
\begin{equation}
\sum_{k\in \Sc} R_k \leq \frac{1}{2}\log \left|I+\frac{P}{2}G(\Sc)
G(\Sc)^T \right|+\frac{1}{2} \min\{|\Sc|,|\Sc^c|\} \log (2|\Sc|) \label{eq:gaussian-ob}
\end{equation}
for all $\Sc\subseteq [1:N]$ with $\Sc^c\cap \Dc \neq \emptyset$. To
show this, first note that the cutset outer bound
\eqref{eq:cutset-mmn} continues to hold with the set of input
distributions satisfying $\E(X_k^2) \le P$, $k \in [1:N]$. For each
$\Sc\subseteq [1:N]$ such that $\Sc^c \cap \Dc \neq\emptyset$, we can
further loosen the cutset outer bound as
\begin{align}
R(\Sc) &\leq I(X(\Sc); Y(\Sc^c) | X(\Sc^c)) \nonumber\\
&=h(Y(\Sc^c)|X(\Sc^c))-h(Y(\Sc^c)|X^N)\nonumber\\
&=h\bigl(G(\Sc)X(\Sc)+Z(\Sc^c)|X(\Sc^c)\bigr)-h(Y(\Sc^c)|X^N)\nonumber\\
&=h\bigl(G(\Sc)X(\Sc)+Z(\Sc^c)\bigr)-h(Y(\Sc^c)|X^N)\nonumber\\
&=\frac{1}{2}\log (2\pi e)^{|\Sc^c|} \Bigl|I+G(\Sc)  K_{X(\Sc)} G(\Sc)^T \Bigr |-\frac{|\Sc^c|}{2}\log (2\pi e)\nonumber\\
&\leq \frac{1}{2}\log \Bigl|I+ \tr(K_{X(\Sc)}) G(\Sc) G(\Sc)^T \Bigr | \label{eq:awgn1}\\
&\leq \frac{1}{2}\log \Bigl|I+ |\Sc| P \cdot G(\Sc) G(\Sc)^T  \Bigr| \label{eq:awgn2}\\
&\leq \frac{1}{2}\log \left|2|\Sc|\cdot I+ 2|\Sc| \frac{P}{2} \cdot G(\Sc) G(\Sc)^T  \right|\nonumber\\
&\leq \frac{1}{2}\log \left|I+ \frac{P}{2} G(\Sc) G(\Sc)^T  \right|+\frac{|\Sc^c|}{2}\log (2|\Sc|), \nonumber
\end{align}
where $K_{X(\Sc)}$ is the covariance matrix of $X(\Sc)$,
\eqref{eq:awgn1} follows since $\tr(K)I-K$ is positive semidefinite
for any covariance matrix $K$~\cite[Theorem 7.7.3]{Horn--Johnson1985},
and \eqref{eq:awgn2} follows since $\tr (K_{X(\Sc)}) \leq |\Sc| P$,
from the power constraint. By rewriting \eqref{eq:awgn2} as
\[
\frac{1}{2}\log \Bigl|I+ |\Sc| P \cdot G(\Sc) G(\Sc)^T  \Bigr|  
=
\frac{1}{2}\log \Bigl|I+ |\Sc| P \cdot G(\Sc)^T G(\Sc)  \Bigr| 
\]
and following similar steps, we also have
\begin{align*}
R(\Sc) &\leq 
\frac{1}{2}\log 
  \left|I+ \frac{P}{2} G(\Sc)^T G(\Sc)  \right|+\frac{|\Sc|}{2}\log (2|\Sc|)\\
&= \frac{1}{2}\log 
  \left|I+ \frac{P}{2} G(\Sc) G(\Sc)^T  \right|+\frac{|\Sc|}{2}\log (2|\Sc|).
\end{align*}

On the other hand, the noisy network coding inner bound in
Theorem~\ref{thm:mmn} yields the inner bound characterized by
the set of inequalities
\begin{equation}
R(\Sc) \le \frac{1}{2}\log \left|I+ \frac{P}{2}G(\Sc)
G(\Sc)^T\right|-\frac{|\Sc|}{2} \label{eq:gaussian-innerbound}
\end{equation}
for all $\Sc\subseteq [1:N]$ with $\Sc^c\cap \Dc \neq \emptyset$. To
show this, first note that by the standard
procedure~\cite{El-Gamal--Kim2010}, Theorem~\ref{thm:mmn} for the
discrete memoryless network can be easily adapted for the AWGN network
with power constraint, which gives the inner bound \eqref{eq:mmn} on
the capacity region with (product) input distributions satisfying
$\E(X_k^2) \le P$, $k \in [1:N]$.

Let $Q= \emptyset$ and $X_k$, $k\in[1:N]$, be i.i.d.\@ Gaussian
with zero mean and variance $P$. Let
\[
\Yh_k=Y_k+\Zh_k,\quad  k\in[1:N],
\] 
where $\Zh_k$, $k\in[1:N]$, are i.i.d. Gaussian with zero mean and
unit variance. Then for each $\Sc\subseteq [1:N]$ such that $\Sc^c\cap
\Dc \neq \emptyset$ and $d\in \Dc$,
 \begin{align*}
 I(\Yh(\Sc); Y(\Sc)|X^N, \Yh(\Sc^c), Y_d) &\leq I(\Yh(\Sc); Y(\Sc)|X^N)\\
 &= h(\Yh(\Sc)|X^N)-h(\Yh(\Sc)|Y(\Sc), X^N)\\
 &= \frac{|\Sc|}{2}\log (4\pi e)-\frac{|\Sc|}{2}\log (2\pi e)\\
 &= \frac{|\Sc|}{2},
 \end{align*}
where the first inequality is due to the Markovity $(\Yh(\Sc^c), Y_d)
\to (X^N, Y(\Sc))\to \Yh(\Sc)$. Furthermore,
\begin{align*}
I(X(\Sc); \Yh(\Sc^c), Y_d | X(\Sc^c)) 
&\geq I(X(\Sc); \Yh(\Sc^c) | X(\Sc^c))  \\
&= h(\Yh(\Sc^c)| X(\Sc^c))- h(\Yh(\Sc^c)| X^N)\\
&= \frac{1}{2}\log(2\pi e)^{|\Sc^c|} \left|2I+ G(\Sc) P G(\Sc)^T\right|-\frac{|\Sc^c|}{2}\log(4\pi e)\\
&= \frac{1}{2}\log \left|I+ \frac{P}{2}G(\Sc) G(\Sc)^T\right|.
\end{align*}
Therefore, by Theorem \ref{thm:mmn}, a rate tuple $(R_1, \ldots, R_N)$ is achievable if 
\[
R(\Sc) 
< \frac{1}{2}\log \left|I+ \frac{P}{2}G(\Sc) G(\Sc)^T\right|-\frac{|\Sc|}{2} 
\]
for all $\Sc\subseteq [1:N]$ such that $\Sc^c \cap \Dc \neq \emptyset$. 

Comparing the cutset outer bound \eqref{eq:gaussian-ob} and inner
bound \eqref{eq:gaussian-innerbound} completes the proof of Theorem 4.

\subsection{AWGN Two-Way Relay Channels} \label{subsec:awgn-twrc}

Recall the model for the AWGN two-way relay
channel~\eqref{eq:awgn-twrc} in Section~\ref{sec:prob-statement}.

Rankov and Wittenben~\cite{Rankov--Wittenben2006} showed that the
amplify--forward (AF) coding scheme results in the inner bound on the
capacity region that consists of all rate pairs $(R_1,R_2)$ such that
\begin{align*}
R_k&< \frac{1}{2}\log\left(\frac{a_k+\sqrt{a_k^2-b_k^2}}{2}\right), \quad k\in\{1,2\}
\end{align*}
for some $\alpha \le \sqrt{P/(g_{13}^2P+g_{23}^2P+1)}$, where
$a_1:=1+\frac{P(g_{12}^2+\alpha^2 g_{32}^2g_{13}^2)}{g_{32}^2\alpha^2+1}$, $a_2:=1+\frac{P(g_{21}^2+\alpha^2 g_{31}^2g_{23}^2)}{g_{31}^2\alpha^2+1}$, $b_1:=\frac{2P\alpha g_{32}g_{13}g_{12}}{g_{32}^2\alpha^2+1}$, and $b_2:=\frac{2P\alpha g_{31}g_{23}g_{21}}{g_{31}^2\alpha^2+1}$.  They
also showed that an extension of the original compress--forward (CF)
coding scheme for the relay channel to the two-way relay channel
results in the following inner bound on the capacity region that
consists of all rate pairs $(R_1,R_2)$ such that
\begin{align*}
R_1&< \C\left(\frac{g_{13}^2 P+(1+\sigma^2)g_{12}^2 P}{1+\sigma^2}\right),\\
R_2&< \C\left(\frac{g_{23}^2 P+(1+\sigma^2)g_{21}^2 P}{1+\sigma^2}\right)
\end{align*}
for some
\[
\sigma^2 \geq 
\max\left\{ \frac{(1+g_{12}^2P)(1+g_{13}^2P)-(g_{12}g_{13}P)^2}{\min\{g_{32}^2, g_{31}^2\}P},\, \frac{(1+g_{21}^2P)(1+g_{23}^2P)-(g_{21}g_{23}P)^2}{\min\{g_{32}^2, g_{31}^2\}P}\right\}.
\]

Specializing Theorem~\ref{thm:dmn-mac} to the two-way relay channel
gives the inner bound that consists of all rate pairs $(R_1,R_2)$
such that
\begin{align*}
R_1 &\le \min\{I(X_1; Y_2, \Yh_3|X_2, X_3, Q),\, I(X_1, X_3; Y_2|X_2, Q)-I(Y_3; \Yh_3|X_1, X_2, X_3, Y_2, Q)\}\\
R_2 &\le \min\{I(X_2; Y_1, \Yh_3|X_1, X_3, Q),\, I(X_2, X_3; Y_1|X_1, Q)-I(Y_3; \Yh_3|X_1, X_2, X_3, Y_1, Q)\}
\end{align*}
for some $p(q)p(x_1|q)p(x_2|q)p(x_3|q)p(\yh_3| y_3, x_3, q)$.  By
setting $Q=\emptyset$ and $\Yh_3=Y_3+\Zh$ with $\Zh \sim
\N(0,\sigma^2)$, this inner bound simplifies to the set of rate pairs
$(R_1, R_2)$ such that
\begin{align}
R_1 &< \min\left\{\C\left(\frac{g_{13}^2 P+(1+\sigma^2)g_{12}^2 P}{1+\sigma^2}\right),\,
\C(g_{12}^2 P+g_{32}^2 P)-\C(1/\sigma^2)\right\},\nonumber\\
R_2 &< \min\left\{\C\left(\frac{g_{23}^2 P+(1+\sigma^2)g_{21}^2 P}{1+\sigma^2}\right),\,
\C(g_{21}^2 P_1+g_{31}^2 P)-\C(1/\sigma^2)\right\} \label{eq:twrc-nnc}
\end{align}
for some $\sigma^2 > 0$.

\subsection{AWGN Interference Relay Channels} 
\label{subsec:awgn-irc}

Recall the model for the AWGN interference relay channel with
orthogonal receiver components in Figure~\ref{fig:irc}.

Djeumou, Belmaga, and Lasaulce~\cite{Djeumou--Belmaga--Lasaulce2009},
and Razaghi and Yu~\cite{Razaghi--Yu2010} showed that an extension of
the original compress--forward (CF) coding scheme for the relay
channel to the two-way relay channel results in the inner bound on the
capacity region that consists of all rate pairs $(R_1,R_2)$ such that
\begin{align*}
R_1 &< \C\left(\frac{(g_{13}^2 + (1+\sigma^2)g_{14}^2)P+(g_{23}g_{14}-g_{24}g_{13})^2P^2}{1+\sigma^2+(g_{23}^2 + (1+\sigma^2)g_{24}^2)P}\right),\\
R_2 &<   \C\left(\frac{(g_{23}^2 + (1+\sigma^2)g_{25}^2)P+(g_{13}g_{25}-g_{15}g_{23})^2P^2}{1+\sigma^2+(g_{13}^2 + (1+\sigma^2)g_{15}^2)P}\right)
\end{align*}
for some 
\begin{align*}
\sigma^2 &\ge \frac{1}{2^{2R_0}-1}\cdot\max\left\{\frac{(g_{13}g_{24}-g_{23}g_{14})^2P^2+a_1}{(g^2_{14}P+g^2_{24}P+1)},\,
\frac{(g_{13}g_{25}-g_{23}g_{15})^2P^2+a_2}{(g^2_{15}P+g^2_{25}P+1)}\right\},
\end{align*}
where
\begin{equation*}
\begin{split}
a_1 &:= (g^2_{13}+g^2_{14})P+(g^2_{23}+g^2_{24})P+1,\\
a_2 &:= (g^2_{13}+g^2_{15})P+(g^2_{23}+g^2_{25})P+1.
\end{split}
\end{equation*}

Razaghi and Yu~\cite{Razaghi--Yu2010} generalized the hash--forward
coding scheme~\cite{Cover--Kim2007, Kim2008a} for the relay channel to
the interference relay channel, in which the relay sends the bin index
(hash) of its noisy observation and destination nodes use list
decoding.  This generalized hash--forward scheme gives the inner bound
on the capacity region that consists of the set of rate pairs $(R_1,
R_2)$ such that
\begin{align*}
R_1 &< \C\left(\frac{g^2_{14}P}{g^2_{24}P+1}\right)+R_0-\C\left(\frac{(g^2_{23}+g^2_{24})P+1}{(g^2_{24}P+1)\sigma^2}\right), \\
R_2 &< \C\left(\frac{g^2_{25}P}{g^2_{15}P+1}\right)+R_0-\C\left(\frac{(g^2_{13}+g^2_{15})P+1}{(g^2_{15}P+1)\sigma^2}\right) 
\end{align*}
for some $\sigma^2 > 0$ satisfying
\begin{align*}
\sigma^2 &\le \frac{1}{2^{2R_0}-1}\cdot\min\left\{\frac{(g_{13}g_{24}-g_{23}g_{14})^2P^2+a_1}{(g^2_{14}P+g^2_{24}P+1)},\,
\frac{(g_{13}g_{25}-g_{23}g_{15})^2P^2+a_2}{(g^2_{15}P+g^2_{25}P+1)}\right\},
\end{align*}
where $a_1$ and $a_2$ are the same as above.  

Specializing Theorem~\ref{thm:dmn-mac} by setting $\Yh_3=Y_3+\Zh$ with
$\Zh \sim \N(0,\sigma^2)$ gives the inner bound that consists of all
rate pairs $(R_1, R_2)$ such that
\begin{align*}
R_1 &< \min\left\{\C(g_{14}^2P)+R_0-\C(1/\sigma^2),\,
 \C\left( \frac{(g^2_{13}+(1+\sigma^2)g^2_{14})P}{1+\sigma^2} \right)\right\},\\
R_2 &< \min\left\{\C(g_{25}^2P)+R_0-\C(1/\sigma^2),\,
\C\left( \frac{(g^2_{23}+(1+\sigma^2)g^2_{25})P}{1+\sigma^2} \right)\right\},\\
R_1+R_2 &< \C((g_{14}^2+g_{24}^2)P)+R_0-\C(1/\sigma^2), \\
R_1+R_2 &< \C\left(\frac{(g_{13}^2+g_{23}^2)P+(1+\sigma^2)(g_{14}^2+g_{24}^2)P+(g_{13}g_{24}-g_{23}g_{14})^2P^2}{1+\sigma^2}\right), \\
R_1+R_2 &< \C((g_{15}^2+g_{25}^2)P)+R_0-\C(1/\sigma^2), \\
R_1+R_2 &< \C\left(\frac{(g_{13}^2+g_{23}^2)P+(1+\sigma^2)(g_{25}^2+g_{15}^2)P+(g_{23}g_{15}-g_{13}g_{25})^2P^2}{1+\sigma^2}\right)
\end{align*}
for some $\sigma^2 > 0$.  By the same choice of $\Yh_3$, the inner
bound in Theorem~\ref{thm:dmn-noise} can be specialized to the set of
rate pairs $(R_1, R_2)$ such that
\allowdisplaybreaks
\begin{align*}
R_1 &< \C\left(\frac{g^2_{14}P}{g^2_{24}P+1}\right)+R_0-\C\left(\frac{(g^2_{23}+g^2_{24})P+1}{(g^2_{24}P+1)\sigma^2}\right), \\
R_1 &< \C\left(\frac{(g_{13}^2 + (1+\sigma^2)g_{14}^2)P+(g_{23}g_{14}-g_{24}g_{13})^2P^2}{1+\sigma^2+(g_{23}^2 + (1+\sigma^2)g_{24}^2)P}\right),\\
R_2 &< \C\left(\frac{g^2_{25}P}{g^2_{15}P+1}\right)+R_0-\C\left(\frac{(g^2_{13}+g^2_{15})P+1}{(g^2_{15}P+1)\sigma^2}\right),\\
R_2 &<   \C\left(\frac{(g_{23}^2 + (1+\sigma^2)g_{25}^2)P+(g_{13}g_{25}-g_{15}g_{23})^2P^2}{1+\sigma^2+(g_{13}^2 + (1+\sigma^2)g_{15}^2)P}\right)
\end{align*}
for some $\sigma^2 > 0$.


\section{Concluding Remarks} 

We presented a new noisy network coding scheme and used it to
establish inner bounds on the capacity region of general
multiple-source noisy networks. This scheme unifies and extends
previous results on network coding and its extensions, and on
compress--forward for the relay channel. We demonstrated that the
noisy network coding scheme can outperform previous network
compress--forward schemes. The reasons are: first, the relays do not
use Wyner--Ziv coding (no binning index to decode), second, the
destinations are not required to decode the compression indices
correctly, and third, simultaneous decoding over all blocks is used.

How good is noisy network coding as a general purpose scheme?  As we
have seen, noisy network coding is optimal in some special cases. It
also performs generally well under high SNR conditions in the network.
In addition, it is a robust and scalable scheme in the sense that the
relay operations do not depend on the specific codebooks used by the
sources and destinations or even the topology of the network.  Noisy
network coding, however, is not always the best strategy. For example,
for a cascade of AWGN channels with low SNR, the optimal strategy is
for the relay to decode the message and then forward it to the final
destination. This simple multi-hop scheme can be improved by using the
information from multiple paths and coherent cooperation as in the
decode--forward scheme for the relay
channel~\cite{Cover--El-Gamal1979} and its extensions to
networks~\cite{Xie--Kumar2005, Kramer--Gastpar--Gupta2005}. Further
improvement can be obtained by only partial decoding of messages at
the relays~\cite{Cover--El-Gamal1979}, and by combining noisy network
coding with partial decode--forward to obtain the type of hybrid
schemes in~\cite{Cover--El-Gamal1979}
and~\cite{Kramer--Gastpar--Gupta2005}.

Another important direction to improve noisy network coding for
multiple sources is to use more sophisticated interference coding
schemes, such as interference alignment~\cite{Cadambe--Jafar2008} and Han--Kobayashi superposition coding~\cite{Han--Kobayashi1981}.

\section*{Acknowledgments}
The authors are grateful to Han-I Su for his comments on earlier
drafts of the paper. The work was supported in part by the DARPA
ITMANET program, NSF CAREER grant CCF-0747111, and MKE/IITA IT R\&D
program 2008-F-004-02.

\appendices
\section{Error Probability Analysis for Theorem~\ref{thm:dmn-mac}}
\label{app:theorem2}

The analysis follows the same steps of the multicast case except that
the union in the third error term of~\eqref{eq:three-error} is over
all $\mv$ such that $(m_k: k\in \Sc_d)\neq (1,\ldots,1)$. Thus,
\begin{align}
&\P(\cup_{\mv}\Ec_\mv)\nonumber\\
&\quad\le \sum_{\mv}\sum_{\lv^b} 
\prod_{j=2}^b \P( \Ac_j(\mv, \lv_{j-1}, \lv_j))\nonumber\\
&\quad\le
\sum_{\substack{\Tc \subseteq [1:N] \\ \Tc\cap \Sc_d \neq \emptyset, d\in \Tc^c}}2^{\sum_{k\in \Tc}nbR_k}\,2^{\sum_{k\neq d} n\Rh_k}2^{(N-1)(b-1)}\, 2^{n\left(- (b-1) \min_{\Sc} (I_1(\Sc) + I_2(\Sc) -
\sum_{k \in \Sc} \Rh_k - \delta(\e))\right)}, \label{eq:errorbnd2}
\end{align} 
where the minimum in \eqref{eq:errorbnd2} is over $\Sc \subseteq [1:N]$ such that $\Tc \subseteq \Sc, d \in \Sc^c$.
Hence, \eqref{eq:errorbnd2} tends to zero as $n \to \infty$ if
\[
R(\Tc) < \frac{b-1}{b} \left(\min_{\substack{\Sc \subseteq [1:N]\\ \Tc \subseteq \Sc, d \in \Sc^c}} \left(I_1(\Sc) + I_2(\Sc) -
\sum_{k\in\Sc} \Rh_k \right)- \delta(\e)\right) - \frac{1}{b} \left(\sum_{k\neq d}
\Rh_k\right)
\]
for all $\Tc \subseteq [1:N]$ such that $\Tc\cap \Sc_d\neq \emptyset$
and $d\in \Tc^c$.  By eliminating $\Rh_k > I(\Yh_k;Y_k|X_k)+\delta(\e')$,
letting $b \to \infty$, and getting rid of inactive inequalities, the
probability of error tends to zero as $n \to \infty$ if
\begin{equation}
R(\Sc) <  I(X(\Sc);\Yh(\Sc^c), Y_d|X(\Sc^c)) - I(\Yh(\Sc); Y(\Sc) | X^N, \Yh(\Sc^c), Y_d) - (N-1)\delta(\e') -\delta(\e), \label{eq:mac-condition}
\end{equation}
for all $\Sc\subseteq [1:N]$ such that $\Sc\cap \Sc_d\neq \emptyset$
and $d\in \Sc^c$. Thus, the probability of decoding error tends to
zero for each destination node $d\in \Dc$ as $n\to \infty$, provided
that the rate tuple satisfies \eqref{eq:mac-condition}. Finally, by
the union of events bound, the probability of error for all
destinations tends to zero as $n \to \infty$ if the rate tuple $(R_1,
\ldots, R_N)$ satisfies
\begin{align*}
R(\Sc)< \min_{d\in \Sc^c\cap \Dc(\Sc)} I(X(\Sc); \Yh (\Sc^c), Y_d|X(\Sc^c)) - I(Y(\Sc);\Yh(\Sc)|&X^{N}, \Yh(\Sc^c), Y_d)
\end{align*}
for all $\Sc\subseteq [1:N]$ such that $\Sc^c\cap \Dc(\Sc) \neq
\emptyset$ for some $\prod_{k=1}^{N} p(x_k) p(\yh_k|y_k,x_k)$. This
completes the proof of Theorem~\ref{thm:dmn-mac} for $Q = \emptyset$.
The proof for the general $Q$ follows by coded time sharing.

\section{Error Probability Analysis for Theorem~\ref{thm:dmn-noise}}
\label{app:theorem3}

Let $M_k$ denote the message sent at node $k\in[1:N]$ and $L_{kj}$,
$k\in [1:N]$, $j\in [1:b]$, denote the index chosen by node $k$ for
block $j$. To bound the probability of error for decoder $d\in \Dc$,
assume without loss of generality that $(M_1,\ldots,M_N) =(1, \ldots,
1)=\mathbf 1$. Define
\begin{align*}
\Ec_0 &:= \bigcup_{j=1}^b\bigcup_{k=1}^N
\bigl\{(\Yvh_{kj}(l_{kj}| L_{k, j-1}), \Uv_{kj}( L_{k, j-1}), \Yv_{kj}) \not \in \aepvar \text{ for all } l_{kj}\bigr\}\\
\Ec_{\mv} &:=
\bigl\{( \{\Xv_{kj}(m_{k}| l_{k,j-1}): k\in \Sc_d\}, \Uv_{1j}(l_{1, j-1}), \ldots,\Uv_{Nj}(l_{N,j-1}), \\
&\qquad\qquad\qquad\Yvh_{1j}(l_{1j}| l_{1,j-1}), \ldots, \Yvh_{Nj}(l_{Nj}| l_{N,j-1}), \Yv_{dj}) \in \aep, j\in [1:b],\\
&\qquad\qquad \qquad \qquad \text{ for some } (\lv_{1}, \ldots, \lv_{b}), \text{ where }  l_{dj}=L_{dj}, j\in[1:b]\bigr\}.
\end{align*}
Here, $\mv:=(m_k:k\in \Sc_d )$ and $\lv_j = (l_{1j}, \ldots, l_{Nj})$
for $j\in[1:b]$.  Then the probability of error is upper bounded as
$\P(\Ec) \leq \P(\Ec_0)+\P(\Ec_0^c \cap \Ec_{\mathbf
  1}^c)+\P(\cup_{\mv \neq \mathbf 1} \Ec_{\mv})$, where $m_d=1$ in
$\mv$. By the covering lemma, $\P(\Ec_0)\to 0$ as $n \to \infty$, if
$\Rh_k > I(\Yh_k;Y_k|U_k)+\delta(\e')$, $k\in [1:N]$, and by the
conditional typicality lemma $\P(\Ec_0^c \cap \Ec_{\mathbf 1}^c )\to
0$ as $n\to \infty$.  For the third term, assume that $\Lv_1=\cdots =
\Lv_b=\mathbf 1$. Define
\begin{align*}
\Ac_{j}(\mv, \lv_{j-1}, \lv_j)
&:= 
\{(\{\Xv_{kj}(m_{k}| l_{k, j-1}): k\in \Sc_d\}, \Uv_{1j}( l_{1, j-1})
,\ldots, \Uv_{Nj}(l_{N, j-1}),\\
&\qquad  \Yvh_{1j}(l_{1j}| l_{1,j-1}), \ldots,
\Yvh_{Nj}(l_{Nj}| l_{N,j-1}),
\Yv_{dj}) \in \aep \}
\end{align*}
for $\mv\neq \mathbf 1$ and all $\lv_j$. 
Then, from similar steps to the multicast case,
\begin{align}
	\P(\Ec_{\mv}) &\le  \sum_{\lv^b}
\prod_{j=2}^b \P( \Ac_j(\mv, \lv_{j-1}, \lv_j)). \nonumber
\end{align}
For each $\lv^b$ and $j \in [2:b]$, let $\Sc_j(\lv^b) \subseteq [1:N]$
such that $\Sc_j(\lv^b)=\{k: l_{k,j-1}\neq 1\}$. We further define
$\Tc(\mv) \subseteq [1:N]$ such that $\Tc(\mv)=\{k: k\in \Sc_d, m_k
\neq 1 \}$. By definition, $d\in \Tc^c(\mv)\cap \Sc^c_j(\lv_{j-1})$,
where $\Tc^c(\mv):= \Sc_d \backslash \Tc(\mv)$.  Then, by the joint
typicality lemma, we can show that
\begin{equation}
\P (\Ac_j(\mv, \lv_{j-1}, \lv_j)) \leq 2^{-n (I_1(\Sc(\lv_{j-1}), \Tc(\mv)) + I_2(\Sc(\lv_{j-1}), \Tc(\mv)) - \delta(\e))}, \nonumber
\end{equation}
where 
\begin{align*}
&I_1(\Sc, \Tc) :=I(X((\Sc \cup \Tc)\cap \Sc_d), U(\Sc); \Yh(\Sc^c), Y_d | X((\Sc^c\cap \Tc^c)\cap \Sc_d), U(\Sc^c)), \text{ and}\\
&I_2(\Sc, \Tc):=\sum_{k \in \Sc} I(\Yh_k; \Yh(\Sc^c\cup\{k'\in\Sc: k' < k\}),Y_d, X(\Sc_d), U^N|U_k).
\end{align*}
Furthermore from the definitions of $\Tc(\mv)$ and $\Sc_j(\lv_{j-1})$, if $\mv \neq \mathbf 1$ with $m_d=1$, then
\begin{align*}
\sum_{\lv_{j-1}}&\,
2^{-n (I_1(\Sc_j(\lv_{j-1}), \Tc(\mv))  + I_2(\Sc_j(\lv_{j-1}), \Tc(\mv))-\delta(\e))}\\
&\le
\sum_{\Sc \subseteq [1:N]: d \in \Sc^c}
2^{-n (I_1(\Sc, \Tc(\mv) )  + I_2(\Sc, \Tc(\mv)) -\sum_{k \in \Sc} \Rh_k- \delta(\e))}\\
&\le
2^{N-1}2^{-n (\min_{\Sc}(I_1(\Sc, \Tc(\mv) )  + I_2(\Sc, \Tc(\mv)) -\sum_{k \in \Sc} \Rh_k- \delta(\e)))}. 
\end{align*}
 Hence,
\begin{align}
\sum_{\mv \ne \mathbf 1}&\sum_{\lv^b} 
\prod_{j=2}^b \P( \Ac_j(\mv, \lv_{j-1}, \lv_j))\nonumber\\
&=
\sum_{\mv \ne \mathbf 1}\sum_{\lv_b}\sum_{\lv^{b-1}} 
\prod_{j=2}^b \P( \Ac_j(\mv, \lv_{j-1}, \lv_j))\nonumber\\
&=
\sum_{\mv \ne \mathbf 1}\sum_{\lv_b} 
\prod_{j=2}^b \sum_{\lv_{j-1}} \P( \Ac_j(\mv, \lv_{j-1}, \lv_j))\nonumber\\
&\le
\sum_{\mv\ne \mathbf 1} \sum_{\lv_b} \prod_{j=2}^{b} \left( \sum_{\lv_{j-1}}
2^{-n (I_1(\Sc_j(\lv_{j-1}), \Tc(\mv))  + I_2(\Sc_j(\lv_{j-1}), \Tc(\mv))-\delta(\e))}
\right) \nonumber\\
&\le
\sum_{\mv\ne \mathbf 1}  \sum_{\lv_b} \prod_{j=2}^{b} \left( 
2^{N-1}2^{-n (\min_{\Sc}(I_1(\Sc, \Tc(\mv) )  + I_2(\Sc, \Tc(\mv)) -\sum_{k \in \Sc} \Rh_k- \delta(\e)))}
\right) \nonumber\\
&\le
\sum_{\substack{\Tc \subseteq \Sc_d \\ \Tc \neq \emptyset, d\in \Tc^c}}2^{\sum_{k\in \Tc}nbR_k}\,2^{\sum_{k\neq d} n\Rh_k}2^{(N-1)(b-1)}\, 2^{n\left(- (b-1) \min_{\Sc} (I_1(\Sc, \Tc) + I_2(\Sc, \Tc) -
\sum_{k \in \Sc} \Rh_k - \delta(\e))\right)}, \label{eq:errorbnd-noise}
\end{align} 
where the minimum in \eqref{eq:errorbnd-noise} is over $\Sc \subseteq [1:N]$ such that $d \in \Sc^c$.
Hence, \eqref{eq:errorbnd-noise} tends to zero as $n \to \infty$ if
\[
R(\Tc) < \frac{b-1}{b} \left(\min_{\Sc \subseteq [1:N], d \in \Sc^c} \left(I_1(\Sc, \Tc) + I_2(\Sc, \Tc) -
\sum_{k\in\Sc} \Rh_k \right)- \delta(\e)\right) - \frac{1}{b} \left(\sum_{k\neq d}
\Rh_k\right)
\]
for all $\Tc \subseteq \Sc_d$ such that $d\in \Tc^c$.
By eliminating $\Rh_k > I(\Yh_k;Y_k|U_k)+\delta(\e')$ and
letting $b \to \infty$, the probability of error tends to zero as $n \to \infty$ if
\[
R(\Tc) < \min_{\Sc \subseteq [1:N], d \in \Sc^c} \left(I_1(\Sc, \Tc) + I_2(\Sc, \Tc) - \sum_{k \in \Sc} I(\Yh_k;Y_k|U_k)
\right)
- (N-1)\delta(\e') - \delta(\e)
\]
for all $\Tc \subseteq \Sc_d$ such that $d\in \Tc^c$.
Finally, note that
\begin{align*}
I_2(\Sc, \Tc) - \sum_{k \in \Sc} I(\Yh_k;Y_k|U_k)
&= 
-\sum_{k \in \Sc}
I(\Yh_k; Y_k | X(\Sc_d), U^N, \Yh(\Sc^c), Y_d, \Yh(\{k'\in\Sc:k' < k\}))\\
&= 
-\sum_{k \in \Sc}
I(\Yh_k; Y(\Sc) | X(\Sc_d), U^N, \Yh(\Sc^c), Y_d, \Yh(\{k'\in\Sc:k' < k\}))\\
&=
-I(\Yh(\Sc); Y(\Sc) | X(\Sc_d), U^N, \Yh(\Sc^c), Y_d).
\end{align*}
Therefore, the probability of error tends to zero as $n \to \infty$ if
\begin{align}
R(\Tc) &< I(X((\Sc\cup \Tc)\cap \Sc_d), U(\Sc);\Yh(\Sc^c), Y_d|X((\Sc^c \cap \Tc^c)\cap \Sc_d), U(\Sc^c)) \nonumber\\
&\qquad - I(\Yh(\Sc); Y(\Sc) | X(\Sc_d), U^N, \Yh(\Sc^c), Y_d) - (N-1)\delta(\e') -\delta(\e) \label{eq:condition1-noise}
\end{align}
for all $\Sc\subseteq [1:N]$ and $\Tc \subseteq\Sc_d$ such that $d\in
\Sc^c$ and $d\in \Tc^c$. Since for every $\Sc\subseteq [1:N]$, $d\in
\Sc^c$ the inequalities corresponding to $\Tc \subsetneq (\Sc \cap
\Sc_d)$ are inactive due to the inequality with $\Tc =\Sc\cap \Sc_d$
in \eqref{eq:condition1-noise}, the set of inequalities can be further
simplified to
\begin{align}
R(\Tc) < I( X(\Tc), &U(\Sc); \Yh (\Sc^c),
Y_{d}| X(\Tc^c), U(\Sc^c))\nonumber\\
 &- I(Y(\Sc);\Yh(\Sc)| X(\Sc_d), U^{N}, \Yh(\Sc^c), Y_{d}) - (N-1)\delta(\e') -\delta(\e) \label{eq:condition2-noise}
\end{align}
for all $\Sc\subseteq [1:N]$ and $\Sc\cap \Sc_d
\subseteq \Tc \subseteq \Sc_d$ such that $d\in \Sc^c$, where $\Tc^c = \Sc_d \backslash \Tc$. Thus, the probability of decoding error tends to zero for each destination
node $d\in \Dc$ as $n\to \infty$, provided that the rate
tuple satisfies \eqref{eq:condition2-noise}. By the union of events bound,
the probability of error tends to zero as $n \to \infty$ if the rate tuple $(R_1, \ldots, R_N)$ satisfies
\begin{align*}
R(\Tc) < & I( X(\Tc), U(\Sc); \Yh (\Sc^c),
Y_{d}| X(\Tc^c), U(\Sc^c)) - I(Y(\Sc);\Yh(\Sc)| X(\Sc_d), U^{N}, \Yh(\Sc^c), Y_{d})
\end{align*}
for all $\Sc\subseteq [1:N]$, $d\in \Dc(\Sc)$, and $\Sc\cap \Sc_d
\subseteq \Tc \subseteq \Sc_d$ such that $\Sc^c\cap \Dc(\Sc) \neq
\emptyset$, where $\Tc^c = \Sc_d \backslash \Tc$ for some
$\prod_{k=1}^{N} p(x_k) p(\yh_k|y_k,x_k)$. This completes the proof of
Theorem~\ref{thm:dmn-noise} for $Q = \emptyset$. The proof for the
general $Q$ follows by coded time sharing.


\section{Comparison to a Previous Extension of the Original Compress--Forward
Scheme} \label{app:kgg} For a DM single-source (node 1) multicast
network with destination nodes $\Dc\subseteq[2:N]$, a hybrid scheme
proposed by Kramer, Gastpar, and Gupta~\cite[Theorem
  3]{Kramer--Gastpar--Gupta2005} gives the capacity lower bound
\begin{equation}
C \ge \max
\min_{d\in \Dc}I(X_1; \hat Y_2^N, Y_d | U_2^N,
X_2^N), \label{eq:kgghyb-mn}
\end{equation}
where the maximum is over
$p(x_1) \prod_{k=2}^{N}p(u_k, x_k)p(\yh_k| u_2^N, x_k,  y_k)$ such that
\begin{align}
&I(\Yh(\Tc); Y(\Tc) | U_2^N, X_2^N, \Yh(\Tc^c), Y_d)+ \sum_{k\in \Tc} I(\Yh_k; X_2^N | U_2^N, X_k) \nonumber \\
&\qquad\leq I(X(\Tc); Y_d | U(\Tc), X(\Tc^c), U_d, X_d) + \sum_{t=1}^T I(U(\Kc_t); Y_{r(t)} | U(\Kc_t^c), X_{r(t)} ) \label{eq:hyb-const}
\end{align}
for all $\Tc \subseteq [2:N]$, all partitions $\{\Kc_t\}_{t=1}^T$ of
$[2:N]$, and all $r(t)\in [2:N]$ such that $r(t)\not \in \Kc_t $. The
complements $\Tc^c$ and $\Kc_t^c$ are the complements of the
respective $\Tc$ and $\Kc_t $ in $[2:N]$.

The hybrid coding scheme achieving lower bound~\eqref{eq:kgghyb-mn}
uses an extension of the original compress--forward scheme for the
relay channel as well as decoding of the compression indices at the
relays. The pure compress--forward scheme without decoding gives the
capacity lower bound
\begin{equation} \label{eq:kgghyb-mn2}
C \ge R^* = \max \min_{d \in \Dc} I(X_1;\hat Y_2^N, Y_d|X_2^N),
\end{equation}
where the maximum is over all pmfs $\prod_{k=1}^N p(x_k)p(y_k|x_k)$ such
that
\[
I(Y(\Tc);\Yh(\Tc)|X_2^N, \Yh(\Tc^c),Y_d) + \sum_{k \in \Tc}I(X_2^N;
\Yh_k | X_k) \le I(X(\Tc);Y_d|X(\Tc ^c), X_d)
\]
for all $\Tc\subseteq [2:N]$ and $\Tc^c=[2:N]\backslash \Tc$.  This is
identical to (32) with $U_j = \emptyset$, $j \in [2:N]$.

In the following we show that the noisy network coding lower bound in
Theorem~\ref{thm:mmn} is uniformly better than lower bound
\eqref{eq:kgghyb-mn2} for every $p(y_2^N|x^N)$. By using similar steps
to those in~\cite[Appendix C]{El-Gamal--Mohseni--Zahedi2006} and some
algebra, lower bound \eqref{eq:kgghyb-mn2} can be upper bounded as
\begin{align}
R^* &\le \max \min_{d\in \Dc} \min_{\Tc\subseteq [2:N]} I(X_1; \hat Y_2^N, Y_d|X_2^N)+I(X(\Tc); Y_d | X(\Tc^c), X_d) \nonumber\\
&\qquad \qquad \qquad \qquad \qquad \qquad \qquad-I(\Yh(\Tc); Y(\Tc) | X_2^N, \Yh(\Tc^c), Y_d)-\sum_{k\in \Tc} I(\Yh_k; X_2^N|X_k) \nonumber\\
&=\max\min_{d\in \Dc} \min_{\Tc\subseteq [2:N]} I(X_1, X(\Tc); \Yh(\Tc^c), Y_d| X(\Tc^c), X_d)-I(\Yh(\Tc); Y(\Tc)| X^N, \Yh(\Tc), Y_d) \nonumber \\
&\qquad \qquad \qquad \qquad\qquad \qquad \qquad-I(X(\Tc); \Yh(\Tc^c)| Y_d, X(\Tc^c), X_d)-\sum_{k\in \Tc} I(\Yh_k; X_2^N| X_k), \label{eq:kggcf-mn2}
\end{align}
where the maximums are over $p(x_1) \prod_{k=2}^{N}p(x_k)p(\yh_k| x_k,
y_k)$.  Here equality \eqref{eq:kggcf-mn2} follows since
\begin{align*}
I(X_1&; \hat Y_2^N, Y_d|X_2^N)+I(X(\Tc); Y_d|X(\Tc^c), X_d)-I(\Yh(\Tc); Y(\Tc) | X_2^N, \Yh(\Tc^c), Y_d)\\
&=
I(X_1; \Yh(\Tc^c), Y_d|X_2^N)+I(X_1; \Yh(\Tc)|X_2^N, \Yh(\Tc^c), Y_d)\\
&\qquad +I(X(\Tc); Y_d|X(\Tc^c), X_d)-I(\Yh(\Tc); Y(\Tc) | X_2^N, \Yh(\Tc^c), Y_d)\\
&=
I(X_1; \Yh(\Tc^c), Y_d|X_2^N)+I(X_1; \Yh(\Tc)|X_2^N, \Yh(\Tc^c), Y_d)+I(X(\Tc); \Yh(\Tc^c), Y_d|X(\Tc^c), X_d)\\
&\qquad
-I(X(\Tc); \Yh(\Tc^c)|X(\Tc^c), Y_d, X_d)-I(\Yh(\Tc); Y(\Tc) | X_2^N, \Yh(\Tc^c), Y_d)\\
&=
I(X_1, X(\Tc); \Yh(\Tc^c), Y_d|X(\Tc^c), X_d)+I(X_1; \Yh(\Tc)|X_2^N, \Yh(\Tc^c), Y_d)\\
&\qquad
-I(X(\Tc); \Yh(\Tc^c)|X(\Tc^c), Y_d, X_d)-I(\Yh(\Tc); Y(\Tc) | X_2^N, \Yh(\Tc^c), Y_d)\\
&=
I(X_1, X(\Tc); \Yh(\Tc^c), Y_d|X(\Tc^c), X_d)+I(X_1, Y(\Tc); \Yh(\Tc)|X_2^N, \Yh(\Tc^c), Y_d)\\
&\qquad
-I(Y(\Tc); \Yh(\Tc)|X_1, X_2^N, \Yh(\Tc^c), Y_d)-I(X(\Tc); \Yh(\Tc^c)|X(\Tc^c), Y_d, X_d)\\
&\qquad-I(\Yh(\Tc); Y(\Tc) | X_2^N, \Yh(\Tc^c), Y_d)\\
&=
I(X_1, X(\Tc); \Yh(\Tc^c), Y_d|X(\Tc^c), X_d)+I(X_1; \Yh(\Tc)|X_2^N, \Yh(\Tc^c), Y(\Tc), Y_d)\\
&\qquad-I(Y(\Tc); \Yh(\Tc)|X_1, X_2^N, \Yh(\Tc^c), Y_d)-I(X(\Tc); \Yh(\Tc^c)|X(\Tc^c), Y_d, X_d)\\
&=
I(X_1, X(\Tc); \Yh(\Tc^c), Y_d|X(\Tc^c), X_d)-I(Y(\Tc); \Yh(\Tc)|X_1, X_2^N, \Yh(\Tc^c), Y_d)\\
&\qquad -I(X(\Tc); \Yh(\Tc^c)|X(\Tc^c), Y_d, X_d)
\end{align*}
for all $\Tc\subseteq [2:N]$, $\Tc^c = [2:N] \backslash \Tc$ and $d\in
\Dc$, where the last equality follows from the Markovity $(X_1,
X(\Tc^c), X_d, \Yh(\Tc^c), Y_d)\to (X(\Tc), Y(\Tc))\to \Yh(\Tc)$.  On
the other hand, Theorem~\ref{thm:mmn} can be simplified by setting $Q
= \emptyset$ and $R_2 = \cdots = R_N = 0$ as
 \begin{align}
C \ge  \max_{\prod_{k=1}^{N}p(x_k)
p(\yh_k|y_k,x_k)} \min_{d\in \Dc}
\min_{\Tc\subseteq [2:N]}
& I( X_1, X(\Tc); \Yh (\Tc^c), Y_d|X(\Tc^c), X_d) \label{eq:cflb-mn}\\
&\qquad \qquad \qquad - I(Y(\Tc);\Yh(\Tc)|X^N, \Yh(\Tc^c), Y_d),  \nonumber 
\end{align}
where $\Tc^c = [2:N] \backslash \Tc$.  Thus, Theorem~\ref{thm:mmn}
achieves a higher rate than \eqref{eq:kgghyb-mn2}
with gap
\[
I(X(\Tc); \Yh(\Tc^c)| Y_d, X(\Tc^c), X_d)+\sum_{k\in \Tc} I(\Yh_k; X_2^N| X_k)
\]
for each $d\in \Dc$ and $\Tc\subseteq [2:N]$.

We now present a simple example for which noisy network coding
performs strictly better than the general hybrid
scheme~\eqref{eq:kgghyb-mn}.  Consider a 4-node noiseless network,
where $\Dc=\{4\}$, $R_2=R_3=R_4=0$, and $Y_2=X_1, Y_3=X_2, Y_4=X_3$
are all binary. From \eqref{eq:noiseless-lb}, we know that that the
noisy network coding lower bound achieves the capacity $C=1$.  On the
other hand, applying \eqref{eq:kgghyb-mn} to the above noiseless
network we get
\begin{align}
I(X_1; \Yh_2, \Yh_3, Y_4| U_2, U_3, X_2, X_3 ) &= I(X_1; \Yh_2, \Yh_3 | U_2, U_3, X_2, X_3) \label{eq:ss1}\nobreak\\
&=I(X_1; \Yh_2 | U_2, U_3, X_2, X_3, \Yh_3) \label{eq:ss2}
\end{align}
where \eqref{eq:ss1} follows from the channel and \eqref{eq:ss2}
follows from the Markovity $X_1 \to (U_2, U_3, X_3, Y_3) \to
\Yh_3$. The constraint~\eqref{eq:hyb-const} corresponding to $\Tc=\{2\}$ and
$r(1)=4$ is
\[
I(\Yh_2; Y_2|U_2, U_3, X_2, X_3, \Yh_3, Y_4)+I(\Yh_2; X_3| U_2, U_3,
X_2)\leq I(X_2; Y_4|U_2, U_3, X_3)+I(U_2; Y_4|U_3),
\]
which can be simplified as
\begin{align*}
I(X_1; \Yh_2 | U_2, U_3, X_2, X_3, \Yh_3)
&\leq I(U_2; X_3|U_3)-I(\Yh_2; X_3|U_2, U_3, X_2)\\
&\leq I(U_2; X_3|U_3) \\
&=0,
\end{align*}
where the equality follows from $U_2 \to U_3 \to X_3$. Thus, the
achievable rate of the hybrid scheme is zero for this particular
example. It can be easily seen that our noisy network coding scheme
outperforms the hybrid scheme for noiseless networks with more than
two relays. Note that in general, due to decoding at the relay nodes,
the hybrid scheme can sometimes perform better than the noisy network
coding scheme without similar augmentation.

\bibliographystyle{IEEEtran}
\bibliography{nit}
\end{document}